\documentclass[letterpaper,12pt]{article}
\pdfoutput=1
\usepackage{jheppub}
\usepackage{amsmath}
\usepackage{graphicx}
\usepackage[font=small]{caption,subcaption}
\usepackage{cancel}
\usepackage[dvipsnames]{xcolor}
\usepackage{color}
\usepackage{mathrsfs}
\usepackage{amssymb}
\usepackage{epstopdf}
\usepackage{dsfont}
\usepackage{float}
\usepackage{amsthm}

\graphicspath{{./figures/}}

  {\list{}{\leftmargin=0.22in\rightmargin=0.3in}\item[]}%
  {\endlist}

\def\be{\begin{equation}}
\def\ee{\end{equation}}

\def\bfchi{\mbox{\boldmath $\chi$}}
\def\bfTheta{\mbox{\boldmath $\Theta$}}
\def\bfSigma{\mbox{\boldmath $\Sigma$}}
\def\bfXi{\mbox{\boldmath $\Xi$}}
\newcommand*{\scri}{\mathscr{I}} 

\addtocontents{toc}{\protect\enlargethispage*{3\baselineskip}} 

\newpage

\title{Infrared Effects in the Late Stages of Black Hole Evaporation}

\author{\'Eanna \'E. Flanagan}
\affiliation{Department of Physics, Cornell University, Ithaca, NY 14853}
\affiliation{Cornell Laboratory for Accelerator-based Sciences and Education (CLASSE), Cornell University, Ithaca, NY 14853}

\emailAdd{eef3@cornell.edu}

\abstract{
As a black hole evaporates, each outgoing Hawking quantum carries
away some of the black holes asymptotic charges associated with the extended Bondi-Metzner-Sachs group.
These include the Poincar\'e charges of energy, linear momentum, intrinsic
angular momentum, and orbital angular momentum or center-of-mass
charge, as well as extensions of these quantities associated with
supertranslations and super-Lorentz transformations, namely
supermomentum, superspin and super center-of-mass charges (also known as soft hair). 
Since each emitted quantum has fluctuations that are of order unity,
fluctuations in the black hole's charges grow over the course of the
evaporation.  We estimate the scale of these fluctuations using a
simple model.  The results are, in Planck units:
(i) The black hole position has a
uncertainty of $\sim M_i^2$ at late times, where $M_i$ is the initial
mass (previously found by Page).
(ii) The black hole mass $M$ has an
uncertainty of order the mass $M$ itself at the epoch when $M \sim
M_i^{2/3}$, well before the Planck scale is reached.
Correspondingly, the time at which the
evaporation ends has an uncertainty
of order $\sim M_i^2$.
(iii) The 
supermomentum and superspin charges are not independent but are determined from the Poincar\'e
charges
and the super center-of-mass charges.
(iv) The supertranslation that characterizes the super center-of-mass charges
has fluctuations at multipole orders $l$ of order unity that that are of order unity in Planck units.
At large $l$, there is a power law spectrum of fluctuations that extends up to $l
\sim M_i^2/M$, beyond which the fluctuations fall off exponentially,
with corresponding total rms shear tensor fluctuations $\sim
M_i M^{-3/2}$.}

\begin{document}
\maketitle

\section{Introduction and summary}
\label{sec:intro}

Hawking's black hole information loss paradox is one of the most enduring
mysteries in theoretical physics: how does information escape from a
black hole during its evaporation? \cite{1976PhRvD..14.2460H,Harlow:2014yka,Marolf:2017jkr}.
Great progress has been made on
this issue in the past few years, using explicit Euclidean path
integral methods.
It is now possible to explicitly compute the Page curve that describes the time evolution of the entanglement
entropy of the emitted Hawking radiation and the black hole, and to show that it is consistent with unitarity  \cite{Almheiri:2019psf,Penington:2019npb,Almheiri:2019hni,Almheiri:2019qdq,Penington:2019kki,Almheiri:2020cfm}.
In addition, the amount of time taken for the information in a diary thrown into a black hole to return in the Hawking radiation
can be reliably computed \cite{Penington:2019npb}.  Nevertheless, some of the processes and computational prescriptions
that arise in the Euclidean domain remain mysterious in the Lorentzian domain, so there is still much to be understood.

A central role in this subject is played by the semiclassical approximation, where
the gravitational field is treated classically (aside from linear perturbations that can be treated
as free gravitons) and the matter fields are treated quantum mechanically.  
This approximation excludes macroscopically large quantum fluctuations in the geometry.
It is the only approximation in which we can compute the full state of the outgoing Hawking radiation.
In addition, the new computational prescriptions for computing the
entanglement entropy of the exact state of the Hawking radiation \cite{Almheiri:2019psf,Penington:2019npb,Almheiri:2019hni,Almheiri:2019qdq,Penington:2019kki,Almheiri:2020cfm} are
expressed in terms of a single semiclassical geometry, as are other
similar powerful theoretical tools and results (the Ryu-Taganacki
formula \cite{Ryu:2006bv,Hubeny:2007xt}, the quantum
focussing conjecture \cite{Bousso:2015mna} and the covariant entropy
bound \cite{Bousso:1999xy}).

On the other hand, it has been known since the work of Page in the
1980s \cite{PhysRevLett.44.301} that the semiclassical approximation
actually fails drastically during the course of black hole
evaporation.  This failure arises as follows: each emitted
quantum carries of a momentum $\sim M^{-1}$ in a random direction,
where $M$ is the mass of the black hole in Planck units, and the corresponding
change in the velocity of the black hole is of order $\sim
M^{-2}$.  This change in velocity causes a net displacement in the center-of-mass of the black hole of order
$\sim M$ after an evaporation time $\sim M^3$.
During the evaporation process we have $n
\sim M^2$ such kicks that accumulate as a random walk, giving a total
net uncertainty in the black hole location of order $\sim \sqrt{n}
M \sim M^2$, much larger than the size of the black hole.
Thus we have superpositions of macroscopically
distinct geometries.

Several authors have argued for the importance of the center-of-mass fluctuations
in understanding the unitarity of black hole evaporation
\cite{2013PhRvD..87h4050N,2013arXiv1308.5686H,Bao:2017who}.
They note that unitarity is required only for 
the evolution in the complete Hilbert space, not in the subspace that corresponds to a single semiclassical geometry,
and that the relative phases of different semiclassical geometries in a
quantum superposition contain information.   However, there are 
counterarguments \cite{2011arXiv1108.0302M,2013JHEP...09..018A} which
suggest that the breakdown of the semiclassical approximation is
fairly innocuous.  First, there are situations where black holes
evaporate in anti-de Sitter space where the center of center-of-mass
spreading is suppressed but where there is still an information loss
paradox \cite{2013JHEP...09..018A}.  Second, the dimension of the
Hilbert space associated with the center-of-mass motion is negligibly
small compared to the relevant scale of the exponential of the
Bekenstein-Hawking entropy, since it scales as a power law in the mass $M$ of the
black hole\footnote{We can impose an infrared cutoff by assuming that the black hole moves on a torus of
  size the evaporation timescale $\sim M^3$, and impose a maximum kinetic energy
of motion of order $\sim 1/M$ [Eqs.\ (\ref{earlypans}) and (\ref{latepansfn1}) below, neglecting logarithmic factors].  This gives
a Hilbert space dimension $\sim M^9$.}.

In Ref.\ \cite{Flanagan:2021ojq} we show that the center-of-mass
fluctuations give rise to large corrections to the angular
distribution of the Hawking radiation.  We also argue there that
those corrections remove one of the primary objections to the proposal that soft hair on
black holes plays a key role in how unitarity of the evaporation is achieved
\cite{Hawking:2016msc,Hawking:2016sgy,Strominger:2017aeh,Pasterski:2020xvn,Cheng:2020vzw},
by increasing the number of soft hair modes that can interact with
outgoing Hawking quanta.

The purpose of this paper is to study the macroscopic fluctuations in
the geometry of evaporating black holes, in more detail than hitherto.
A more detailed understanding may be useful for eventually extending
some of the theoretical tools discussed above to situations where
infrared quantum fluctuations are large.  It may also shed light on
the role of soft hair. Finally, some of the results derived here were
used in the computations of Ref.\ \cite{Flanagan:2021ojq}.

The theoretical framework we use to study these fluctuations is as follows.
In the classical theory, the geometry of a stationary black hole is determined
by the conserved charges on future null infinity, including the soft
hair charges associated with extensions of the Bondi-Metzner-Sachs (BMS)
group \cite{Hawking:2016msc,Hawking:2016sgy}. We assume that the this
property remains true in the quantum theory, and
evolve the charges using a simple model, described in Secs.\ \ref{sec:Newtonian1} and \ref{sec:pres} below.
We extend
similar previous studies
\cite{PhysRevLett.44.301,2013PhRvD..87h4050N} of
the fluctuations in a number of directions:

\begin{itemize}

\item We extend the computations to late times when the black hole
  mass $M$ is small compared to its initial mass $M_i$, by making use
  of the approximation that the fluctuations in charges are small
  compared to their expected values [Eqs.\ (\ref{lateansfn1}) below].  This approximation is valid
  until $M \sim \sqrt{M_i}$.  The variance in the center of mass
  location does not evolve significantly as the black hole shrinks
  from $M \sim M_i$ to $M \ll M_i$, but remains $\sim M_i^2$.
  
\item We compute the evolution of the fluctuations in the mass of the
  black hole.  This is of order unity in Planck units after an evaporation times, but
  grows at late times according to $\Delta M \sim M_i^2/M^2$ [Eq.\ (\ref{latemansfn1}) below].
  It follows that $\Delta M \sim M$ when $M \sim M_i^{2/3}$.
At this epoch, there is an order unity amplitude for the evaporation to be completed ($M =
0$), but there is also an order unity amplitude for the black
hole mass to be macroscopic, $M \sim M_i^{2/3}$.

\item We extend previous studies to include the charges associated
with an extension of the BMS group \cite{Strominger:2017zoo,Campiglia:2014yka,CL,Cnew}.   
These charges are reviewed in Sec.\ \ref{sec:reviewbms} below, and
are higher-$l$ analogs of the center-of-mass, momentum and spin that are
encoded in the asymptotic metric near future null infinity.
We show that for evaporating black holes only some of these charges
are independent. The independent charges can be taken to be the
so called super center-of-mass charges, or soft hair.  These charges
can be parameterized in terms of the supertranslation required to set
them to zero, a function $\Phi$ on the two-sphere with dimensions of
length with only $l \ge 2$ components.
See
Secs.\ \ref{sec:reviewbms}, \ref{sec:statregions} and Appendix \ref{app:indepcharges}
for more details.

The accumulated fluctuations in the supertranslation $\Phi$ are
relatively small.  For multipole orders $l$ of order unity, the
fluctuations are of order unity in Planck units [Sec.\ \ref{sec:results1}].  There is also a
contribution to the fluctuations associated with
quanta that have only partially arrived at future null infinity by the
time the charges are measured, which we compute in Sec.\ \ref{sec:transient}.
This contribution gives a power law spectrum of fluctuations extending up to $l \sim M_i^2/M$
which is subdominant at early times, but becomes dominant when $M$
becomes small compared to $M_i^{2/3}$ [Sec.\ \ref{sec:transient1}].

\end{itemize}

The organization of this paper is as follows.  The predictions of our model for the evolution of the Poincar\'e charges
are given in Sec.\ \ref{sec:Newtonian}.  Section \ref{sec:bms} extends
this model to include soft hair charges.  In Sec.\ \ref{sec:transient} we consider some additional contributions
to the fluctuations of the soft hair charges that are associated with
quanta that have only partially arrived at future null infinity by the
time the charges are measured.
Preliminary versions of some
of the results were presented at the conferences
\cite{2006APS..APRB11002F,EF2016}.

\section{Simple stochastic model for evolution of black hole
  Poincar\'e charges}
\label{sec:Newtonian}

\subsection{Definition of the evaporation model}
\label{sec:Newtonian1}

In this section we define a stochastic process that gives a crude model
of the evaporation of a black hole, including fluctuations in its
Poincar\'e conserved charges.  We believe that it captures the
dominant effects of the fluctuations.
A variant of this model was first introduced by Page \cite{PhysRevLett.44.301} and explored
in more detail by Nomura, Varela and Weinberg \cite{2013PhRvD..87h4050N}.
Page estimated the fluctuations in position of the black
hole after an evaporation time.  Below we extend his computations to
late times and also estimate the mass fluctuations of the black hole.
A more sophisticated model which includes
all the BMS and extended BMS charges will be given in Sec.\ \ref{sec:bms} below.  The results
from that more sophisticated model for the Poincar\'e charges agree qualitatively with those of
the simple Page model discussed here.

We start by describing the motivation for the model.
Consider the evaporation of a black hole of mass $M \gg 1$, in Planck
units with $8 \pi G = \hbar = c = 1$.  Roughly one Hawking quanta per time
$\Delta t \sim M$ is emitted; this is explicit in an orthonormal wavepacket mode basis.
Each quantum carries an energy $\Delta E\sim M^{-1}$
and spatial momentum $\Delta p \sim M^{-1}$, which change the energy
and momentum of the black hole by corresponding amounts.
The fractional fluctuations in $\Delta E$ and $\Delta p$ are of order
unity, since they are carried by a single quantum.  Also the spatial
momentum can be in a random direction.

The classical stochastic process is defined by
\begin{subequations}
\label{simple}
\begin{eqnarray}
\label{massevolve}
M_{n+1} &=& M_n - \frac{\epsilon_n}{M_n}, \\
\label{timeevolve}
t_{n+1} &=& t_n + M_n,\\
\label{momevolve}
p_{n+1} &=& p_n - \frac{\epsilon_n \delta_n}{M_n}, \\
\label{posnevolve}
x_{n+1} &=& x_n + p_n.
\end{eqnarray}
\end{subequations}
Here $n = 1,2,3 \ldots$ labels the steps, one for each emitted quantum.
The variable $M_n$ is the mass of the black hole at step $n$.
The quantity $\epsilon_n$ is a random variable which takes on the
values $0$ and $1$ with probability each of $1/2$.
The mass evolution equation (\ref{massevolve}) describes the mass of
the black hole being reduced, with a probability to occur of order unity,
by each emitted quantum.  The emission process takes a time of order the
current mass of the black hole, as encoded in the time evolution
equation (\ref{timeevolve}); $t_n$ is the time of step $n$.
The spatial momentum of the black hole after step $n$ is $p_n$, and
its evolution is governed by Eq.\ (\ref{momevolve}).
The magnitude of the emitted spatial momentum is $\epsilon_n/M_n$, the same as
the energy, since the Hawking quantum is massless.  However the
momentum can be in either direction (we are using a one dimensional
model of the black hole motion).  The directionality is encoded in the
random variable $\delta_n$, which takes on the
values $-1$ and $1$ with equal probability\footnote{
The interpretation of individual steps in the model (\ref{simple}) as
individual quanta should not be taken too literally.  In particular, if we quantize a
  scalar field near future null infinity on a spherical harmonic
  basis, then the operator (\ref{Delta4P}) below that describes linear
  momentum radiated consists entirely of cross terms between different
  modes, rather than individual modes carrying linear momentum.
  However,
  consider the following slight generalization of the model: at each
  timestep there are two independent random variables $\epsilon_n$ and
  $\epsilon_n'$ that take on the values $0$ and $1$ with equal
  probability, with $\epsilon_n=1$ representing the emission of a
  $l=0$ scalar quantum and $\epsilon_n'=1$ representing an $l=1$
  quantum.  Then Eqs.\ (\ref{simple}) effectively hold with the
  $\epsilon_n$ in Eq.\ (\ref{massevolve}) replaced by $\epsilon_n + \epsilon_n'$
and with the $\epsilon_n$ in Eq.\ (\ref{momevolve}) replaced by $\epsilon_n \epsilon_n'$.
The qualitative predictions of the model are unchanged by this refinement.}.
The change in the position $x_n$ of the black hole during step $n$ is
the momentum $p_n$ divided by the mass $M_n$ times the time interval $M_n$,
which yields the evolution equation (\ref{posnevolve}).
All of the variables $\epsilon_n$, $\delta_n$ for $n = 1,2,3 \ldots$
are uncorrelated.  The initial conditions at $n=1$ are taken to be
$M_1 = M_i$, the initial black hole mass,
and $p_1 = x_1 = t_1 = 0$.

The evaporation model (\ref{simple}) clearly incorporates a number of
simplifications and approximations.  However we believe that the key
predictions of the model are robust and are insensitive to these
simplifications. Some of the approximations are:

\begin{itemize}

\item It incorporates only one spatial dimension,
and treats only the Poincar\'e conserved charges of the black hole,
neglecting the additional charges associated with the BMS algebra
and its extensions.
These restrictions will be lifted to some extent in Sec.\ \ref{sec:bms} below.


\item It treats all the fluctuations classically rather than quantum
  mechanically.  This restriction will be lifted to some extent in Sec.\ \ref{sec:beyond} below.

\item It models a continuous process as a discrete process.
  However we believe that this idealization does not affect the
  scale of the late time fluctuations predicted by the model.

\item It neglects any initial fluctuations in the black hole's
  conserved charges.  This is acceptable since at late times
the fluctuations will be dominated by the cumulative effects of the emitted
Hawking quanta, for any reasonable estimate of initial
fluctuations\footnote{For example, for a particle of mass $M_i$, the
  standard quantum limit
on the uncertainty in position
after a time $t$ is $\Delta x \gtrsim \sqrt{ t/M_i}$
\cite{Braginsky:1992:QM:171674}.  This uncertainty is of
order $\sim M_i$ after an evaporation time $t \sim M_i^3$, much smaller
than the late time uncertainty $\Delta x \sim M_i^2$ due to the
Hawking quanta, cf.\ Eq.\ (\ref{latexansfn1}).}.

\item Clearly, the the model could be generalized and made more precise
by inserting dimensionless parameters of order unity
into each of the equations (\ref{simple}); this would not change the
qualitative predictions.

\item The motion of the black hole is treated non-relativistically.
This is consistent since the motion is still non-relativistic at the time
the model breaks down, cf.\ Eq.\ (\ref{latepansfn1}) below.

\end{itemize}

The key feature of the model is that independent, uncorrelated
fluctuations are introduced into the black holes conserved charges at
each timestep or each emission event.  Those fluctuations ultimately
originate in incoming modes of quantum fields at past null infinity,
which are orthonormal and so uncorrelated for the incoming vacuum state.

\subsection{Early time predictions: large position fluctuations}
\label{sec:earlytime}

At times small compared to an evaporation time, $t \ll M_i^3$,
simple random walk arguments can be used to estimate the fluctuations
in the black hole charges, as first done by Page
\cite{PhysRevLett.44.301}
and explored
in more detail by Nomura, Varela and Weinberg \cite{2013PhRvD..87h4050N}.
In this section we review these early time predictions of the model.

We define $\tau$ to be the time since black hole formation in units of
the evaporation time,
$
\tau = t/M_i^3.
$
The fluctuations in the various quantities are
\begin{subequations}
\label{earlyans}
\begin{eqnarray}
\label{earlyxans}
\Delta x &=&  \frac{M_i^2 \tau^{3/2}}{\sqrt{6}} \left[ 1 + O(\tau) + O(\tau^{-1}
  M_i^{-2}) \right],\\
\label{earlypans}
\Delta p &=&  \sqrt{\tau/2} \left[1 + O(\tau)  \right], \\
\label{earlymans}
\Delta M &=&  \sqrt{\tau}/2 \left[1 + O(\tau)\right].
\end{eqnarray}
\end{subequations}
Thus, after an evaporation time, the fluctuations in the mass and
momentum are of
order unity in Planck units, whereas the fluctuations in position are
large, of order $\sim M_i^2$.

To derive these estimates we use the approximation $M_n = M_i$ on the
right hand sides of Eqs.\ (\ref{simple}); this is valid up to
fractional corrections of order $(M_i - M_n) / M_i \sim \tau$ [Eq.\
(\ref{bartans}) below].
Solving the momentum evolution equation (\ref{momevolve}), squaring and taking an
expectation value and using
\be
\langle \epsilon_n \epsilon_m \rangle = \frac{1}{4}(1 + \delta_{nm}), \ \ \ \ 
\langle \delta_n \delta_m \rangle = \delta_{nm},
\label{vareps}
\ee
gives $\langle p_n^2 \rangle = (n-1)/(2 M_i^2)$.  Combining this with $t_n
= (n-1) M_i$ from Eq.\ (\ref{timeevolve}) yields the momentum fluctuation
estimate (\ref{earlypans}).  Similarly the mass evolution equation
(\ref{massevolve}) in
this approximation yields
$
M_n = M_i - \sum_{j=1}^n\epsilon_j/M_i,
$
and an analogous argument yields the mass fluctuation estimate
(\ref{earlymans}).
Finally solving the position evolution equation (\ref{posnevolve})
yields
$
x_{n+1} = \sum_{j=1}^n \sum_{k=1}^j \epsilon_k \delta_k/M_i.
$
Squaring and taking the expectation value gives
$
\langle x_n^2 \rangle = n^3/(6 M_i^2),
$
up to fractional corrections $\sim 1/n$,
which yields the estimate (\ref{earlyxans}).\footnote{Using similar
  methods one can show that $\langle x p \rangle = \sqrt{3/4} \Delta x
  \, \Delta p$ at leading order, so the position and momentum fluctuations are somewhat
  correlated as one might expect.}

\subsection{Late time predictions: large mass fluctuations}
\label{sec:latetime}
We now extend the computations of the previous subsection
from the regime $\tau \ll 1$ to $\tau \sim 1$, by approximating the
black hole mass fluctuations\footnote{Here we mean the mass
fluctuations at fixed $n$ given by Eq.\ (\ref{latemansfn}) below,
not the much larger fluctuations at fixed time $t$ given by Eq.\ (\ref{latemans}).}
to be small compared to the
expected value of the mass.  We will show that this approximation
remains valid until $M \sim \sqrt{M_i}$.

The results for the fluctuations at a time when the expected black hole mass is
$M$ are derived in Appendix \ref{app:derive} and are
\begin{subequations}
\label{lateans}
\begin{eqnarray}
\label{latexans}
\Delta x &=&
\bigg\{
\frac{1}{4} (M_i^2 - M^2) (M_i^2 - 3 M^2)
+ M^4 \ln \left(\frac{M_i}{M}\right)\bigg\}^{1/2}
\nonumber \\ && \times
\left[1 + O\left(
\frac{\sqrt{M_i^2-M^2}}{M^2}\right)\right],\\
\label{latepans}
\Delta p &=&  \sqrt{ \ln \left(\frac{M_i}{M}\right)}
\left[1 + O\left(\frac{1}{M^2} \right)+
O\left(
\frac{M_i^2-M^2}{M^4}\right)\right], \\
\label{latemans}
\Delta M &=&
\bigg\{ \frac{3 M_i^4 - 8 M_i^3 M + 12 M_i^2 M^2 - 7 M^4}{24 M^4}
\bigg\}^{1/2}
\left[1 + O\left(
\frac{\sqrt{M_i^2-M^2}}{M^2}\right)\right].
\end{eqnarray}
\end{subequations}
These results are consistent with Eqs.\ (\ref{earlyans}) above in
their common domain of validity $\tau \ll 1$, using the relation $M^3
= M_i^3 (1 - 3 \tau/2)$.

The results simplify in the late time limit when the black hole is small, $M \ll M_i$, to
\begin{subequations}
\label{lateansfn1}
\begin{eqnarray}
\label{latexansfn1}
\Delta x &=&
\frac{1}{2}M_i^2
\left[1 +
O\left(\frac{M}{M_i}\right)+
O\left(
\frac{M_i}{M^2}\right)\right],\\
\label{latepansfn1}
\Delta p &=&  \sqrt{ \ln \left(\frac{M_i}{M}\right)}
\left[1 + O\left(
\frac{M_i^2}{M^4}\right)\right], \\
\label{latemansfn1}
\Delta M &=&  \frac{M_i^2}{\sqrt{8} M^2}
\left[1 +
O\left(\frac{M}{M_i}\right)+
O\left(
\frac{M_i}{M^2}\right)\right].
\end{eqnarray}
\end{subequations}
The key qualitatively new feature in this regime
is the enhanced fluctuations of the black hole mass, which become of
order the mass itself, $\Delta M \sim M$, at
$M \sim M_i^{2/3}$.  As discussed in the introduction, this implies that there
is an order unity amplitude for the evaporation to be completed at the
same time as there is an order unity amplitude for the black  
hole mass to be macroscopic, $M \sim M_i^{2/3}$.

There is an elementary argument for the large, late time mass
fluctuations, which is as follows \cite{2007IJTP...46.2204H}.
Consider a black hole with initial mass $M_i$.  After an evaporation
time $\sim M_i^3$, when the mass is $M_i/2$, the spread in mass is of
order unity, from the
random walk estimates of Sec.\ \ref{sec:earlytime}.
Consider now two different histories, one with mass $M_i/2$ at this
time and one with mass $M_i/2+1$.
For the subsequent evolution of these two histories, we 
consider just the evolution of the mean mass for simplicity, neglecting
fluctuations produced by the subsequent Hawking emission.
The corresponding mean masses a time $t$
later are $[(M_i/2)^3-3t/2]^{1/3}$ and $[(M_i/2+1)^3-3t/2]^{1/3}$, and when the mass
of the first history is zero, the mass of the second history is $\sim M_i^{2/3}$.
Thus the overall fluctuations in mass at this time must be at least $\sim M_i^{2/3}$.

\subsection{Numerical simulation of model}

\begin{figure}[h]
  \begin{center}
    \captionsetup{width=5in}
    \includegraphics[width=5in]{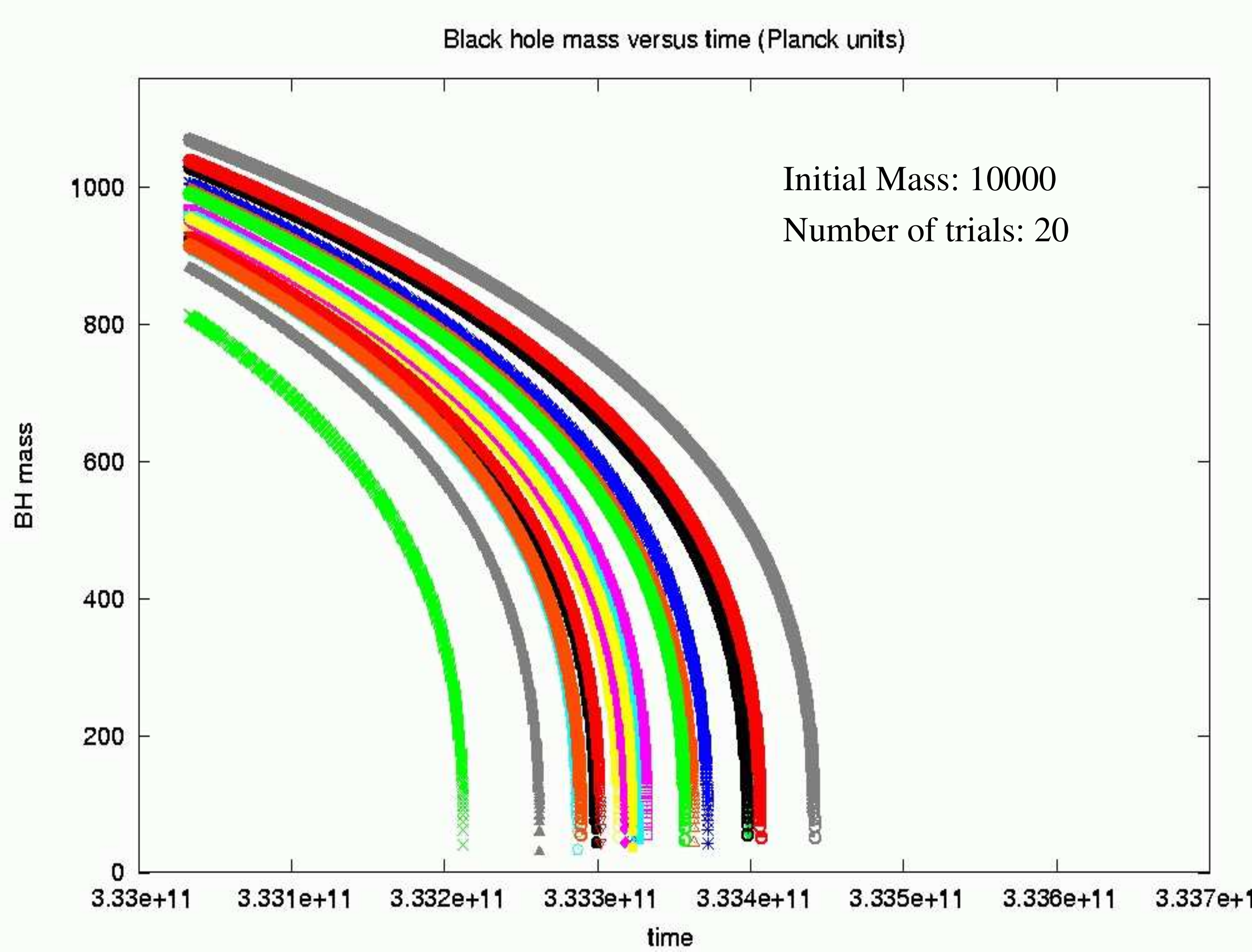}
\caption{Black hole evaporation histories showing mass as a function
  of time in Planck units, obtained by numerically integrating
Eqs.\ (\ref{simple}).  The simulation confirms the late time
prediction $\Delta M \sim M \sim M_i^{2/3}$ at the endpoint of
evaporation.}
\label{fig-st}
    \end{center}
\end{figure}

It is also straightforward to numerically simulate the stochastic
model (\ref{simple}) of the black hole evaporation.
Representative results are shown in Fig.\ \ref{fig-st}, for an
initial mass of $M_i=10^4$ in Planck units, and showing 20 independent
trials.  The results confirm the analytic predictions of
a spread of $\Delta t \sim M_i^2$ in the endpoint of evaporation,
and a spread $\Delta M \sim M \sim M_i^{2/3}$ in mass at the endpoint.
Although the analytic calculations break down once $\Delta M \sim M$,
the numerical results indicate that the fluctuations do not
dramatically change after this occurs.

\subsection{Beyond the classical stochastic model}
\label{sec:beyond}

So far, the motion of the black hole has been treated classically.
However, generalizing to a more detailed quantum mechanical treatment does not
qualitatively change the results, as we now outline.
The black hole center-of-mass motion can be described by its Wigner
function ${\cal W}({\bf p}, {\bf x})$,
with variance-covariance matrix $\Sigma^{AB} = \langle \langle \zeta^A \zeta^B
\rangle \rangle - \langle \langle \zeta^A \rangle \rangle \langle
\langle \zeta^B \rangle \rangle$ with $\zeta^A = ({\bf p}, {\bf x})$
and $\langle \langle f \rangle \rangle \equiv \int d^3 p \int d^3 x f
{\cal W}$ for any function $f$.
As before, we idealize the evolution as a series of $n$ steps, each of
which has two parts.  First, the black hole emits a quantum, under
which by momentum conservation $\bfSigma$ transforms as
\be
\bfSigma \to \bfSigma +\left( \begin{array}{cc}
\sigma_\gamma^2 \delta_{ij} & 0  \\
0 & 0  \end{array} \right),
\ee
where $\sigma_\gamma \sim M^{-1}$ is the uncertainty in momentum of the emitted Hawking quantum.
Second, the black hole evolves freely for a time $\Delta t \sim M$, under which $\bfSigma \to {\bf R} \cdot \bfSigma \cdot {\bf R}^T$, where
\be
{\bf R} = \left( \begin{array}{cc}
 \delta_{ij} & 0  \\
 \delta_{ij} \Delta t/M & \delta_{ij}  \end{array} \right).
\ee
After $n \sim M^2$ steps, the effect of the initial value of $\bfSigma$ is negligible, and the
predicted scalings of position uncertainty and momentum uncertainty agree with Eqs.\ (\ref{earlyans}) above.

\section{Evolution and fluctuations of black hole extended Bondi-Metzner-Sachs charges}
\label{sec:bms}

In this section we generalize the Newtonian model of the previous section to include
all of the Bondi-Metzner-Sachs (BMS) charges of the black hole
 and in addition the charges associated with
the extended BMS algebra
\cite{Barnich:2009se,Barnich:2011ct,2011JHEP...12..105B,Strominger:2014pwa,FN}.

As discussed in the introduction, in this paper we focus on charges of the black hole measured
at future null infinity.  One could instead consider the symmetries and charges defined on the black hole horizon,
which has a different symmetry algebra
\cite{Hawking:2016sgy,Hawking:2016msc,Chandrasekaran:2018aop,Donnay:2016ejv,Donnay:2015abr,Eling:2016xlx,Cai:2016idg,Carlip:2017xne,Blau:2015nee,Penna:2017bdn,Grumiller:2018scv}.
These charges are related to charges at future null infinity by global
conservation laws, as detailed in Ref.\ \cite{Hawking:2016msc},
assuming that one can find the appropriate identification between
horizon symmetry generators and asymptotic symmetry generators
(currently known in some special cases).  However our approach here
follows the perspective of a distant asymptotic observer.

\subsection{Review of BMS and extended BMS charges}
\label{sec:reviewbms}

We start by reviewing the nature of the BMS and extended BMS charges of asymptotically flat spacetimes.
For more details on this topic see the review by Strominger
\cite{Strominger:2017zoo} and the expositions \cite{Cnew,Compere:2019gft}.
The BMS group is the group of asymptotic Killing vectors that act on
spacetimes which are asymptotically flat at future null infinity \cite{1962RSPSA.269...21B,
1962RSPSA.270..103S,1962PhRv..128.2851S}.
Associated with each
asymptotic Killing vector ${\vec \xi}$ or generator of the group, and
with each cut of future null infinity, there is a conserved charge
$Q({\vec \xi})$ \cite{1984CQGra...1...15D,Wald:1999wa}.

We follow the notation of Flanagan and Nichols (FN) \cite{FN}; the
notation (FN,2.1) will mean Eq.\ (2.1) of FN.
In retarded Bondi coordinates $(u,r,\theta^A) = (u,r,\theta^1,\theta^2)$,
the metric of an asymptotically flat spacetime near future null infinity can be written as
\cite{Barnich:2009se,Barnich:2010eb,Strominger:2013jfa,He:2014laa,Kapec:2014opa,Strominger:2014pwa,Pasterski:2015tva}.
\begin{eqnarray}
ds^2 &=& - \left[1  - \frac{2m }{r} + O \left( \frac{1}{r^2} \right)
  \right]  du^2 - 2 \left[ 1+ O\left(\frac{1}{r^2} \right) \right] du dr \nonumber \\
&&+ r^2 \left[ h_{AB} + \frac{1}{r} C_{AB} + O \left( \frac{1}{r^2}
  \right) \right] (d\theta^A -
{\cal U}^A du) (d\theta^B - {\cal U}^B du),
\label{metric}
\end{eqnarray}
where
\be
{\cal U}^A = - \frac{1}{2 r^2} D_B C^{AB} + \frac{1}{r^3} \bigg[ - \frac{2}{3} N^A
+ \frac{1}{16} D^A(C_{BC} C^{BC})
+ \frac{1}{2} C^{AB} D^C C_{BC} \bigg]
+ O(r^{-4}),
\label{NAdef}
\ee
$A,B = 1,2$, and all the functions that appear in the metric are
are functions of $u$ and $\theta^A$.  The metric $h_{AB}$ is the unit
round metric on the two-sphere and is used to raise and lower capital
Roman indices, and $D_A$ is the associated covariant derivative.
There are three important, leading-order functions in the metric's expansion coefficients
\cite{1962RSPSA.269...21B,Barnich:2009se,Barnich:2010eb,Strominger:2013jfa,He:2014laa,Kapec:2014opa,Strominger:2014pwa,Pasterski:2015tva}:
the Bondi mass aspect $m(u,\theta^A)$, the angular-momentum
aspect $N^A(u,\theta^A)$, and the symmetric tensor
$C_{AB}(u,\theta^A)$ whose derivative
\be
N_{AB} = \partial_u C_{AB}
\label{news}
\ee
is the Bondi news tensor.
Evolution equations for these
metric functions in terms of retarded time are given by (FN,2.4),
(FN,2.11a) and (FN,2.11b).
The leading order components of the stress energy tensor are given by
(FN,2.6) and involve functions ${\hat T}_{uu}(u,\theta^A)$ and ${\hat T}_{uA}(u,\theta^A)$.

The BMS algebra is the algebra of infinitesimal diffeomorphisms on
null infinity that map from one Bondi frame $(u,\theta^A)$ to
another.  A general BMS generator can be written as [Eq.\ (FN,2.13)]
\begin{align}
{\vec \xi} = \left[ \alpha(\theta^A) + \frac{1}{2} u D_A Y^A(\theta^B)
\right] \partial_u + Y^A(\theta_B) \partial_A,
\label{bmsinfinitesimal}
\end{align}
where $Y^A$ is a globally smooth conformal Killing vector on the 2-sphere (the set of which
is isomorphic to the Lorentz algebra), and
$\alpha$ is an arbitrary smooth function that parameterizes the
supertranslation transformations.

Two different extensions of the BMS algebra have been proposed.
Barnich and Troessaert \cite{Barnich:2009se,Barnich:2010eb} suggested
an extension that includes
all local conformal Killing fields $Y^A$ on a \(2\)-sphere, allowing
isolated singular points.  This replaces the Lorentz algebra
of vector fields with an infinite dimensional Virasoro algebra; see
Ref.\ \cite{Strominger:2017zoo} for more details.
Campiglia and Laddha \cite{Campiglia:2014yka,CL} suggested extending the Lorentz transformations to include all
smooth infinitesimal diffeomorphisms $Y^A$ on a \(2\)-sphere.

We focus here on the second extension, whose status can be summarized
as follows.  It arises as the symmetry group of an extended phase
space of general relativity at future null infinity, in which
fewer\footnote{The key idea justifying the extension is that one should only fix diffeomorphism
  degrees of freedom that correspond to degeneracy directions of the
  presymplectic form, and the standard construction of the BMS algebra
  fixes some degrees of freedom that are not degeneracy directions.} of
the diffeomorphism degrees of freedom are fixed than is usual
\cite{Campiglia:2014yka,CL,Flanagan:2019vbl}.
While the presymplectic current of general relativity diverges at null
infinity in the extended phase space, by
exploiting a redefinition freedom\cite{Wald:1999wa} in the presymplectic current it can be
made finite.  Comp\`ere, Fiorucci and Ruzziconi use this method to
construct a finite presymplectic current and derive charges
$Q({\vec \xi})$ associated with each symmetry generator ${\vec \xi}$ in the
extended algebra \cite{Cnew}.  However, their construction uses a
specific coordinate system, and so is not obviously local and covariant
(which would be necessary for uniqueness).  Indeed it can be shown there is no redefinition
of the presymplectic current that is local and covariant throughout the
spacetime and which makes the presymplectic current finite at null
infinity \cite{Flanagan:2019vbl}.  Nevertheless, the charges defined by
Ref.\ \cite{Cnew} can be shown indirectly to be covariant and unique \cite{Compere:2019bua}.
They have also been shown to be consistent with the leading and
subleading soft graviton theorems \cite{Cnew}.

The charges associated with a symmetry of the form
(\ref{bmsinfinitesimal}), with $Y^A$ an arbitrary smooth vector field on the two sphere, on a
cut $u = $ constant of a stationary region of future null infinity can be written as \cite{Cnew}
\be
Q({\vec \xi}) = \frac{1}{8 \pi} \int d^2 \Omega \left[ 2 \alpha m +
  Y^A {\hat N}_A \right],
\label{charge-general}
\ee
where\footnote{The formula (\ref{hatNdef}) corrects Eq.\ (3.5) of FN, which is valid only for BMS
  symmetries and for even parity extended symmetries, since the correction terms (fifth and sixth terms) in Eq.\ (\ref{hatNdef}) are parity odd.
We will apply this formula however only in the even parity case for which the BMS formula would suffice.}
\begin{eqnarray}
{\hat N}_A &=& N_A - u D_A m - \frac{1}{16} D_A (C_{BC} C^{BC})
- \frac{1}{4} C_{AB} D_C C^{BC}
-\frac{u}{4} D_B D_A D_C C^{BC}
\nonumber \\ &&
+ \frac{u}{4} D_B D^B D^C C_{CA}.
\label{hatNdef}
\end{eqnarray}

In a given Bondi frame\footnote{Note that the various charges discussed here
  mix together under transformations of Bondi frame, see, for example,
  Appendix B of FN.  In this paper we adopt the convention of using the initial Bondi
  frame, associated with the stationary state to which the black hole settles
down after it is first formed.}, the charges we consider are
(Sec.\ III of FN):  

\begin{itemize}

\item  The {\it Bondi four momentum} $P^\alpha$ which is encoded in $l
  = 0,1$ pieces of the the Bondi mass aspect $m(u,\theta^A)$
  and conjugate to normal translations.

\item  The {\it supermomentum} charges which are encoded in the $l \ge
  2$ pieces of $m(u,\theta^A)$ and are conjugate to supertranslations.  They
  encode a separate energy conservation law at each angle
  \cite{Strominger:2013jfa}. 

\item  The {\it angular momentum} $J^{\alpha\beta}$ which is encoded in the
$l=1$ piece of ${\hat N}_A(u,\theta^A)$ and is conjugate to the
  Lorentz generators $Y^A$ (conformal Killing vectors). As usual this
  can be split into intrinsic angular momentum, and orbital angular
  momentum or center-of-mass charge (center-of-mass minus velocity
  times time).

\item  The {\it superspin} charges which are encoded in the
 magnetic parity piece of the $l \ge 2$ piece of ${\hat N}_A(u,\theta^A)$,
 and are conjugate to $l \ge 2$ magnetic parity symmetry generators $Y^A$ in the
 extended algebra.
 They encode a separate 
 conservation law for intrinsic angular momentum at each angle \cite{Pasterski:2015tva}.

\item  The {\it super center-of-mass} charges which are encoded in the
  electric parity piece of the $l \ge 2$ piece of ${\hat N}_A$,
  and are conjugate to $l \ge 2$ electric parity symmetry generators $Y^A$ in the
extended algebra.  
They encode a separate conservation
law for orbital angular momentum or center-of-mass charge
at each angle \cite{FN,Nichols:2018qac}. 
In the context of black holes they are also called {\it soft hair} \cite{Strominger:2014pwa,Hawking:2016msc,Hawking:2016sgy}.

We can parameterize these charges in terms of a set of symmetric,
tracefree tensors ${\cal J}_{ij}$, ${\cal J}_{ijk}, \ldots$ as
follows.  For any symmetric, tracefree Cartesian tensor $
{\cal Y}^{i_1 \ldots i_l}$, we consider the symmetry generator
(\ref{bmsinfinitesimal}) with $\alpha = 0$ and with $Y_A = D_A ({\cal
  Y}^L n_L)$, where $L$ is the multi-index $i_1 \ldots i_l$, $n_L =
n_{i_1} \ldots n_{i_l}$, and $n_i$ is the unit vector $(\sin \theta
\cos \varphi, \sin \theta \sin \varphi, \cos \theta)$.  We define the symmetric tracefree tensor
${\cal J}_L$ by demanding that the corresponding charge
(\ref{charge-general}) is ${\cal Y}^L {\cal J}_L$.  The tensor ${\cal
  J}_L$ is related to the $l$th multipole electric parity piece of ${\hat N}_A$ by
\be
   {\hat
     N}^{\rm el}_A = g_l D_A( {\cal J}_L n^L)
\label{eq:stt}
   \ee
   with $g_l =  2 (2 l + 1)!! / (l
(l+1) l!)$, from Eq.\ (\ref{charge-general}) and using the identity
(C5) of Ref.\ \cite{Compere:2019gft}.

\end{itemize}


\subsection{Charges in stationary regions of future null infinity}
\label{sec:statregions}

We will idealize the evaporation of a black hole a sequence of
transitions between stationary states: after each Hawking quantum is
emitted, the black hole settles down to a stationary state, then
the next quantum is emitted, and so on.  

In regions of future null infinity that are stationary, the various
charges discussed above are not all independent.  This follows from
the fact that there exists a canonical Bondi frame associated with the
stationary region in which the metric functions take a simple form
that encode the mass and intrinsic spin (see, for example, Sec. II.D
of FN):
\begin{subequations}
\label{stat}
\begin{eqnarray}
\label{mstationary}
m(\theta^A) &=& m_0 = {\rm constant}, \\
\label{CABstationary}
C_{AB}(\theta^A) &=& 0, \\
N_A(\theta^A) &=& {\rm magnetic\ parity,\ }l=1.
\end{eqnarray}
\end{subequations}
Now a general Bondi frame will be related to
the canonical frame (\ref{stat}) by a nonlinear BMS transformation of
the form (FN,2.12), parameterized by a Lorentz transformation and a
supertranslation, and therefore the metric
functions $m$, $C_{AB}$ and $N_A$ in the general frame
encode just one infinite family of charges, and not three (see
Appendix \ref{app:indepcharges}).
We will focus here on the independent charges, which we take to be the
momentum $P^\alpha$, angular momentum $J^{\alpha\beta}$, and super
center-of-mass charges; the other charges can be determined from
these.

We also show in Appendix \ref{app:indepcharges} that the super center-of-mass charges are determined
to a good approximation by the shear tensor $C_{AB}$, so we focus on
this quantity in subsequent sections rather than on $N_A$.   Specifically, 
we decompose $C_{AB}$ in terms of
an electric parity potential $\Phi$ and a magnetic parity potential $\Psi$ via
\be
C_{AB} = D_A D_B \Phi - h_{AB} D^2  \Phi/2 +
\epsilon_{C(A} D_{B)} D^C \Psi,
\label{Phidef}
\ee
where we take the $l=0,1$ pieces of $\Phi$ and $\Psi$ to vanish.  We
also expand the electric parity potential $\Phi$ in terms of a set of
symmetric, tracefree tensors
\be
\Phi = \sum_{l \ge 2} \sum_{i_1 \ldots i_l} {\cal Q}_{i_1 \ldots i_l} n^{i_1} \ldots
n^{i_l} = \sum_{l \ge 2} \sum_L {\cal Q}_L n^L.
\label{Phiexpand}
\ee
The relation between the tensors ${\cal Q}_L$ and the super
center-of-mass charges ${\cal J}_L$ is given in
Eqs.\ (\ref{JQrelation2}) and (\ref{JQrelation}) of Appendix
\ref{app:indepcharges}.

\subsection{Stationary to stationary transitions and changes in the charges}
\label{sec:transition}

In our model of black hole evaporation, each emission of a Hawking quantum is idealized as a stationary
to stationary transition as viewed at future null infinity.  Specifically, this
means that the spacetime at some early retarded time $u_1$ is vacuum near
future null infinity, and is also approximately stationary there.
There
is subsequently a burst of gravitational waves and/or matter energy
flux to infinity, and the spacetime is again vacuum and
approximately stationary near future null infinity at some later
retarded time $u_2$.  In this section we will give formulae for the changes in
the BMS and extended BMS charges in such transitions, in terms of fluxes to null infinity
of mass-energy or gravitational-wave energy.  These formulae will be
one foundation of our model of black hole evaporation of
Sec.\ \ref{sec:evolution} below, and are derived in Appendix \ref{app:fluxes}.

The changes in the linear momentum $P^\alpha$ and angular momentum
$J^{\alpha\beta}$ have the same form as in special relativity, but
with the stress-energy fluxes supplemented by gravitational wave
terms.  The total energy radiated per unit solid angle in
either matter or gravitational waves is
\be
\Delta {\cal E} =  \int_{u_1}^{u_2} du \left[  {\hat T}_{uu} +
  {\cal T}_{uu} \right],
\label{DeltaE}
\ee
where ${\hat T}_{uu} = \lim_{r \to \infty} r^2 T_{uu}$, ${\cal T}_{uu} = N_{AB} N^{AB}/(32 \pi)$ and $N_{AB}$ is the
news tensor (\ref{news}).  The Bondi 4-momentum is
given by
$P^\alpha = (E,{\bf P}) =\int d^2 \Omega (1,{\bf n}) m/ (4 \pi)$,
from Eqs.\ (FN,3.5), (FN,3.7) and (FN,3.9),
and the change in Bondi
4-momentum is
\be
\Delta P^\alpha = (\Delta E,
\Delta {\bf P}) = - \int d^2 \Omega (1,{\bf n}) \Delta {\cal
  E}.
\label{Delta4P}
\ee
Similarly we define 
\be
\Delta {\cal E}_A =  \int_{u_1}^{u_2} du \left[  {\hat T}_{uA} +
  {\cal T}_{uA} \right],
\label{DeltaEA}
\ee
where ${\hat T}_{uA} = \lim_{r \to \infty} r^2 T_{uA}$ and
${\cal T}_{uA}$ 
is a kind of gravitational wave angular momentum flux
given in terms of $C_{AB}$ and $N_{AB}$ by Eq.\ (FN,3.23).
The quantity $\Delta {\cal E}_A$ can be interpreted as angular momentum
radiated per unit solid angle in either matter or gravitational waves.
We also define the quantity 
\be
{\widetilde {\Delta {\cal E}}} =  \int_{u_1}^{u_2} du \, u \left[  {\hat T}_{uu} +
  {\cal T}_{uu} \right].
\label{tildeDeltaE}
\ee
The components of angular momentum are given in terms of ${\hat N}_A$
by
\begin{subequations}
\begin{eqnarray}
\label{Jijdef}
    J_{ij} &=& \frac{1}{4\pi} \int d^2 \Omega e^A_{\ [i} n_{j]} {\hat N}_A, \\
\label{J0idef}
    J_{0i} & =& \frac{1}{8\pi} \int d^2 \Omega \, e^A_{i}  {\hat N}_A,
\end{eqnarray}
\end{subequations}
from Eqs.\ (\ref{charge-general}), (FN,3.5), (FN,3.8) and (FN,3.9),
where $e^i_A= D_A n^i$.
The changes in these quantities are (see Appendix \ref{app:fluxes})
\begin{subequations}
\label{DeltaJs}
\begin{eqnarray}
\label{DeltaJij}
\Delta J_{ij} &=& -2 \int d^2 \Omega e^A_{\ [i}
  n_{j]} \Delta {\cal E}_A, \\
\label{DeltaJ0i}
\Delta J_{0i} &=& - \int d^2 \Omega \bigg[ e^A_{\ i}
 \Delta {\cal E}_A
-  n_i {\widetilde {\Delta {\cal E}}} \bigg].
\end{eqnarray}
\end{subequations}

Finally we turn to the super center-of-mass charges.
As discussed in Sec.\ \ref{sec:statregions} above, the center-of-mass
charges are encoded in the electric parity potential $\Phi$ for the
shear tensor $C_{AB}$ defined in Eq.\ (\ref{Phidef}).
The change $\Delta \Phi$ in this potential (which encodes the gravitational wave memory)
is given by\footnote{The formula (\ref{DeltaPhi}) is valid to quadratic order in the
radiated momentum $-\Delta {\bf P}$, see Ref.\ \cite{Satishchandran:2019pyc} for an
exact version.}
\be
{\cal D} \Delta \Phi = 4 \pi {\cal P}  \Delta {\cal E} + \frac{6 n_i
  n_j}{m_0} \left[ \Delta P^i \Delta P^j -
  \frac{1}{3} \Delta {\bf P}^2 \delta^{ij} \right].
\label{DeltaPhi}
\ee
Here $m_0$ is the rest mass of the initial Bondi 4-momentum, $\Delta {\bf P}$ is the momentum change (\ref{Delta4P}), 
${\cal P}$ is the projection operator that sets to zero the $l=0,1$ pieces of functions on the
sphere, and ${\cal D}$ is the angular differential operator 
\be
   {\cal D} = D^2/4 + D^4/8
   \label{calDdef}
   \ee
where $D^2 = D_A D^A$.
The formula (\ref{DeltaPhi}) is valid in initial rest frames, i.e., 
Bondi frames in which the spatial components of the initial Bondi 4-momentum vanish.

\subsection{Evolution model}
\label{sec:evolution}

We now describe the evolution model for the black hole evaporation process, which
is based on the same philosophy as the simple Newtonian model of Sec.\
\ref{sec:Newtonian} above.  As before the evaporation is treated as a series of discrete steps, and
each step is a classical stochastic event.  The only generalization is that
all of the BMS charges are included instead of just the
Poincar\'e charges.

After the black hole is first formed, with initial mass $M_i$, it
rapidly settles down to a stationary state, on a timescale $\sim M_i$.
We will call the canonical Bondi frame associated with this initial
stationary state
the {\it initial Bondi frame}.
After each Hawking quantum is emitted, we assume that the black hole settles down
again to a new stationary state, with a new associated Bondi frame
which we call the {\it instantaneous Bondi frame}, before the next quantum is emitted.
The changes in the BMS charges, that is, the charges carried off by the Hawking quantum, will have a simple
universal form in the instantaneous Bondi frame, but
their form in the initial Bondi frame
will be more complicated and will depend on the values
of all the BMS charges.

We now describe the model in more detail.  We denote by $x^\alpha = (u,r,\theta^A)$
the initial Bondi frame.  At the $n$th step, the black hole BMS
charges
 in this frame are
4-momentum $P_n^\alpha$, angular
momentum $J_n^{\alpha\beta}$, and the tensor $C^n_{AB}(\theta^A)$
which encodes the super center-of-mass charges.
Our goal is to derive a formula for the charges at step $n+1$
in the initial Bondi frame, in terms of the corresponding values at the $n$th step, and
also the changes in the charges in the instantaneous Bondi frame.

We compute the BMS transformation from the initial frame to the $n$th instantaneous frame in two stages.
First, we make a supertranslation parameterized by a function $\beta$
on the 2-sphere (see Appendix B of FN),
to a Bondi frame $x^{\hat \alpha} = ({\hat u}, {\hat r}, \theta^{\hat A})$.
We can write this function as $\beta = t^0 - t^i n_i + \beta_2$, where $n_i = (\sin\theta \cos\varphi, \sin\theta \sin \varphi ,\cos\theta)$, $t^\mu$ is a 4-vector associated with the normal translation piece of the transformation, and $\beta_2$
is purely $l \ge 2$.  The charges in the hatted Bondi frame are [Eqs.\ (FN,B7) and (FN,B8)]
\begin{subequations}
\begin{eqnarray}
P_n^{\hat \alpha} &=& P_n^\alpha, \\
J_n^{{\hat \alpha}{\hat\beta}} &=& J_n^{\alpha\beta} - t_n^\alpha P_n^\beta + t_n^\beta P_n^\alpha + \delta J^{\alpha\beta}[\beta_2], \\
C^n_{{\hat A}{\hat B}} &=& C^n_{AB} - 2 D_A D_B \beta_2 + h_{AB} D^2 \beta_2.
\label{Cnhat}
\end{eqnarray}
\end{subequations}
Here $\delta J^{\alpha\beta}[\beta_2]$ is given by Eqs.\ (FN,B7) with $\beta$ replaced by $\beta_2$.
Next, we choose the supertranslation $\beta_2$ to make
\be
C^n_{{\hat A}{\hat B}}=0,
\label{Cnhat0}
\ee
which determines $\beta_2$ uniquely as shown in Sec.\ II.D of FN.  We
also choose the translation to make
\be
P_{n\,{\hat \alpha}} J_n^{{\hat \alpha}{\hat\beta}} = 0,
\ee
which makes the hatted frame be a center-of-mass frame.
A translation which achieves this is
\be
t_n^\beta = \frac{1}{M_n^2} P_{n\,\alpha}( J_n^{\alpha\beta} + \delta
J^{\alpha\beta}[\beta_2]),
\ee
where $M_n^2 = - {\vec P}_n^2$.

Next, we perform a boost $\Lambda_{\ \,{\hat \alpha}}^{\bar \alpha}$ from the hatted Bondi frame to the
instantaneous Bondi frame $x^{\bar \alpha} = ({\bar u}, {\bar r},
\theta^{\bar A})$.  From Eqs.\ (FN,B3) and (FN,B6) the charges
transform as
\begin{subequations}
\begin{eqnarray}
P_n^{{\bar \alpha}} &=& \Lambda_{\ \, {\hat \alpha}}^{\bar
  \alpha}  P_n^{{\hat
    \alpha}}, \\
J_n^{{\bar \alpha}{\bar \beta}} &=& \Lambda_{\ \, {\hat \alpha}}^{\bar
  \alpha} \Lambda_{\ \, {\hat \beta}}^{\bar \beta} J_n^{{\hat
    \alpha}{\hat \beta}}, \\
C^n_{{\bar A}{\bar B}} &=& \omega_\varphi \, \varphi_* C^n_{{\hat
    A}{\hat B}} =0.
\end{eqnarray}
\end{subequations}
Here $\varphi : S^2 \to S^2$ is the conformal isometry of the two
sphere associated with the boost, as described after Eq.\ (FN,B3),
and $\omega_\varphi$ is given by Eq.\ (\ref{omegaformula}).
The boost is determined in the usual way by the requirement that the new frame be
a rest frame, ie
\be
P_n^{{\bar i}} =0.
\label{whichboost}
\ee

Finally, in the instantaneous Bondi frame, the changes in the charges
due to the emission of a Hawking quantum are
\begin{subequations}
\begin{eqnarray}
P_{n+1}^{{\bar \alpha}} &=&
P_{n}^{{\bar \alpha}} + \Delta P_{n}^{{\bar \alpha}},
 \\
J_{n+1}^{{\bar \alpha}{\bar \beta}} &=&
J_{n}^{{\bar \alpha}{\bar \beta}} + \Delta J_{n}^{{\bar \alpha}{\bar \beta}},
\\
C^{n+1}_{{\bar A}{\bar B}} &=&
C^{n}_{{\bar A}{\bar B}} +\Delta C^{n}_{{\bar A}{\bar B}}.
\end{eqnarray}
\end{subequations}
The prescription we use for the changes $\Delta P_{n}^{{\bar \alpha}}$,
$\Delta J_{n}^{{\bar \alpha}{\bar \beta}}$ and $\Delta C^{n}_{{\bar A}{\bar B}}$
is discussed in Sec.\ \ref{sec:pres} below.

We now transform the new charges back to initial Bondi frame, and
express the results in terms of the initial Bondi frame components of
the old charges.  The final result is\footnote{We have replaced ${\hat
    \alpha}$ with $\alpha$ in the
  indices of the Lorentz transformation.  This is a slight notational
  inconsistency but there is no real inconsistency because
  the Bondi frames $x^{\hat \alpha}$ and $x^\alpha$ differ only by a supertranslation.}
\begin{subequations}
\label{esystem}
\begin{eqnarray}
\label{p1}
P_{n+1}^{\alpha} &=&
P_{n}^{\alpha} + \Lambda_{\bar \alpha}^{\ \,\alpha} \Delta P_n^{\bar \alpha},
 \\
\label{j1}
J_{n+1}^{\alpha\beta} &=&
J_{n}^{\alpha\beta} +
\Lambda_{\bar \alpha}^{\ \,\alpha} \Lambda_{\bar \beta}^{\ \,\beta}
\Delta J_{n}^{{\bar \alpha}{\bar \beta}}
+ \frac{2}{M_n^2} P_{n\,\gamma} \left( J_n^{\gamma[\alpha} + \delta
  J^{\gamma[\alpha} \right) \Lambda_{\bar \beta}^{\ \,\beta]} \Delta
P_n^{\bar \beta},\ \ \
\\
\label{c1}
C^{n+1}_{AB} &=&
C^{n}_{AB} + \omega_{\varphi^{-1}} \varphi^{-1}_* \Delta C^{n}_{{\bar A}{\bar B}}.
\end{eqnarray}
\end{subequations}
Here the right hand sides are functions of
the changes in the charges in the instantaneous Bondi frame, and of
the charges at step $n$:
the Lorentz transformation $\Lambda_{\bar \alpha}^{\ \,\alpha}$
 and associated map $\varphi$ are
determined as a function of $P_n^\alpha$ by Eq.\ (\ref{whichboost}),
while the quantity $\delta J^{\alpha\beta}$ is given as a function of
$C^n_{AB}$ by Eqs.\ (\ref{Cnhat}), (\ref{Cnhat0}) and Eq.\ (FN,B7) with $\beta$
replaced by $\beta_2$.

From the structure of these evolution equations we see that the
evolution (\ref{p1}) of the 4-momentum is uncoupled from that of the angular
momentum and super center-of-mass.  In particular, this implies that
the results of the simple model of Sec.\ \ref{sec:Newtonian} above for the
4-momentum evolution should still be valid.  We also see that
the super center-of-mass evolution (\ref{c1}) is uncoupled from the angular
momentum, and can be computed once the 4-momentum evolution is known.
Finally, the angular momentum evolution (\ref{j1}) depends on both the 4-momentum
evolution and the super center-of-mass evolution.

\subsection{Slow motion approximation}
\label{sec:slow}

The velocity of the black hole is of order $\sim 1/M$ at late times,
up to a logarithmic factor, from Eq.\ (\ref{latepansfn1}) above.
This is small compared to unity until the Planck scale $M \sim 1$; in
particular it is small compared to unity when $M \sim M_i^{2/3}$, the
epoch when $\Delta M \sim M$.  Therefore in the evolution equations
(\ref{esystem}) it is a good approximation to treat the velocity as
small.  Expanding to linear order in the velocity ${\bf v}_n$ and
splitting into space and time components and writing ${\vec P}_n = (E_n,
{\bf P}_n) = M_n (1,{\bf v}_n)$
yields
\begin{subequations}
  \label{eq:sys}
\begin{eqnarray}
\label{mn1}
M_{n+1} &=& M_n + \Delta E_n +  v^i_n \Delta P^i_n, \\
\label{pn1}
P^i_{n+1}& =& P^i_n + \Delta P^i_n + v^i_n \Delta
E_n,\\
\label{je}
J_{n+1}^{0i} &=& \left( 1 + \frac{\Delta E_n}{M_n} \right) J_n^{0i} +
\Delta J_n^{0i}
+ \frac{\Delta E_n}{M_n} J_n^{ij} v_n^j
+ \frac{1}{M_n} (v_n^j \Delta P_n^j) J_n^{0i}
\nonumber \\ &&
- \frac{1}{M_n} (v_n^j  J_n^{0j}) \Delta P_n^i
- \frac{8}{5} \Delta E_n {\cal Q}^n_{ij} v_n^j, \\
J_{n+1}^{ij} &=& J_n^{ij} + \Delta J_n^{ij}
- \frac{2}{M_n} J_n^{0[i} \Delta P_n^{j]}
- 2 \Delta J_n^{0[i} v_n^{j]}
-2 \frac{\Delta E_n}{M_n} J_n^{0[i} v_n^{j]}
\nonumber \\ &&
+ \frac{2}{M_n} v_n^k J_n^{k[i} \Delta P_n^{j]}
+\frac{16}{5} v_n^k
{\cal Q}^{n\,k[i}  \Delta P_{n}^{j]}, \\
C^{n+1}_{AB} &=& C^n_{AB} + \Delta C^n_{AB} + {\cal L}_{\vec Y}
\Delta C^n_{AB}
- \frac{1}{2} \Delta C^n_{AB} D_C Y^C ,
\end{eqnarray}
\end{subequations}
where $Y^A = - v_n^i e^A_{\ i}$.
Here we have parameterized the electric quadrupole piece of $C^n_{AB}$
in terms of a tensor ${\cal Q}^n_{ij}$ using the definitions (\ref{Phidef})
and (\ref{Phiexpand}).
Equations (\ref{eq:sys}) show explicitly the leading coupling of the super
center-of-mass to the angular momentum evolution, which occurs through
the quadrupole ${\cal Q}^n_{ij}$.

The velocity terms in these equations are suppressed relative to the
other terms by a factor $\sim M$, so we can take the zero velocity
limit.
This yields
\begin{subequations}
\label{kkk}
\begin{eqnarray}
\label{e11}
M_{n+1} &=& M_n + \Delta E_n, \\
\label{p11}
P^i_{n+1}& =& P^i_n + \Delta P^i_n,\\
\label{jj1}
J_{n+1}^{0i} &=& \left(1 + \frac{\Delta E_n}{M_n} \right) J_n^{0i} + 
\Delta J_n^{0i}, \\
\label{jj2}
J_{n+1}^{ij} &=& J_n^{ij} + \Delta J_n^{ij}
- \frac{2}{M_n} J_n^{0[i} \Delta P_n^{j]}, \\
C^{n+1}_{AB} &=& C^n_{AB} + \Delta C^n_{AB} .
\label{cnans}
\end{eqnarray}
\end{subequations}

The angular momentum evolution is more transparent if we switch to a
different set of variables, namely the center of mass $X_n^i = (P_n^i
u_n - J_n^{0i})/M_n$ and the intrinsic angular momentum $S_n^{ij} =
J_n^{ij} - 2 X_n^{[i} P_n^{j]}$.  Here $u_n$ is the value of the
retarded time coordinate $u$ along future null infinity $\scri^+$ at step $n$.
We also note from Eq.\ (\ref{DeltaJ0i}) that for $u$ large compared to
$\Delta u_n = u_{n+1}-u_n \sim M$, the change in the space-time
components of the angular momentum can be written as
\be
\Delta J_n^{0i} = {\overline {\Delta J}_n^{0i}} + u_n \Delta P_n^i,
\ee
where ${\overline {\Delta J}_n^{0i}}$ is given by the first term on
the right hand side of Eq.\ (\ref{DeltaJ0i}).
Using these new variables to rewrite Eqs.\ (\ref{jj1}) and (\ref{jj2}) 
and making use of Eqs.\ (\ref{mn1}) and (\ref{pn1}),
we obtain our final result for the evolution prescription in the slow motion approximation:
\begin{subequations}
\label{ss24}
\begin{eqnarray}
\label{e111}
M_{n+1} &=& M_n + \Delta E_n, \\
\label{p111}
P^i_{n+1}& =& P^i_n + \Delta P^i_n,\\
\label{posnevolve1}
X_{n+1}^i &=& X_n^i +
\frac{\Delta u_n}{M_n} P_n^i - \frac{{\overline {\Delta J}_n^{0i}}}{M_n}, \ \ \ \ \\
\label{spinevolve}
S_{n+1}^{ij} &=& S_n^{ij} + \Delta J_n^{ij} + 2 \frac{\Delta E_n}{M_n} X_n^{[i} P_n^{j]}
 + \frac{2}{M_n} {\overline {\Delta J}_n^{0[i}} P_n^{j]},\\
C^{n+1}_{AB} &=& C^n_{AB} + \Delta C^n_{AB} .
\label{cnans1}
\end{eqnarray}
\end{subequations}

\subsection{Changes in the charges in the instantaneous Bondi frame}
\label{sec:pres}

To complete the model we need to give a prescription for the changes
$\Delta E_n$, $\Delta {\bf P}_n$, $\Delta J_n^{ij}$, ${\overline {\Delta J}_n^{0i}}$
and $\Delta C^n_{AB}$ to the charges in the $n$th instantaneous Bondi frame.
For the first four of these charges, we can use simple order of magnitude estimates
as we did in the Newtonian model of Sec.\ \ref{sec:Newtonian} above.
However, for the super center-of-mass charges,
the relevant physics is less familiar, which is why we derived
general formulae for the changes in the charges in terms of fluxes to infinity
in Sec.\ \ref{sec:transition} above.  We now use those formulae as a guide to develop a prescription.

For a single Hawking quantum, Eqs.\ (\ref{DeltaE}) and (\ref{Delta4P}) are consistent with a radiated energy
of order $\sim M^{-1}$ and a radiated linear momentum of order $\sim
M^{-1}$ in a time $\Delta u \sim M$ if the flux is of order
\be
{\hat T}_{uu} + {\cal T}_{uu} \sim M^{-2}.
\label{efsc}
\ee
Similarly, the angular momentum $\Delta J^{ij}$ carried by a single
quantum should be of order unity\footnote{The linear momentum is of
  order $\sim M^{-1}$ and the effective displacement from the center
  of mass of the black hole is at most of order $\sim M$.}.  This
implies from Eq.\ (\ref{DeltaJij})
that the angular momentum flux should scale as\footnote{These
  scalings are also consistent with a scalar field model of the
  outgoing Hawking flux: For the outgoing solution $\Phi =
  f(u,\theta^A)/r + O(r^{-2})$ we have ${\hat T}_{uu} = f_{,u}^2$ and
  ${\hat T}_{uA} = f_{,u} f_{,A}$, with $f \sim 1$, $\partial_u \sim
  M^{-1}$, $\partial_A \sim 1$.}
\be
{\hat T}_{uA} + {\cal T}_{uA} \sim M^{-1},
\ee
and so we expect the first line in Eq.\ (\ref{DeltaJ0i}) to scale as
${\overline {\Delta J}_n^{0i}} \sim 1$.
A simple model for the changes in the Poincar\'e charges,
consistent with these estimates and along the lines of the Newtonian
model of Sec.\ \ref{sec:Newtonian}, is given by
\begin{subequations}
\label{bmsmodel}
\begin{eqnarray}
\Delta E_n &=& - \frac{\epsilon_n}{M_n}, \\
\Delta P^i_n &=&  \frac{\epsilon_n}{M_n} \Xi_n^i,\\
\Delta J_n^{ij} &=& \epsilon_n \epsilon^{ijk} \chi_n^k, \\
{\overline {\Delta J}_n^{0i}} &=& \epsilon_n \Theta_n^i.
\end{eqnarray}
\end{subequations}
Here as before $\epsilon_n = 0$ or $1$ is a random variable,
and $\bfXi_n$, $\bfchi_n$ and $\bfTheta_n$ are independently and randomly distributed on
the unit sphere.

For the super-center-of-mass charges, we parameterize $\Delta C^n_{AB}$
in terms of a potential $\Delta \Phi_n$ as in Eq.\ (\ref{Phidef}), which
in turn is given by Eq.\ (\ref{DeltaPhi}).  We can neglect the second
term on the right hand side, since it is smaller than the first term
by a factor $\sim M^2$.  The remaining term is the $l \ge 2$ piece of
the time integral of the energy flux, which by Eqs.\ (\ref{DeltaE})
and (\ref{efsc}) scales as $\sim M^{-1}$.  We therefore adopt the simple
model\footnote{In a more complete treatment, the quantum field would
be split up into independent modes, and the stress energy components
would be given by expressions quadratic in the field.  The
nonlinearities would then induce correlations between the $l=0,1$ piece
of the energy flux and the $l\ge 2$ pieces.  Those correlations are
not present here, which is a shortcoming of our simple model.
However, the correlations are unlikely to dramatically reduce the
final fluctuations in the BMS charges.}
\be
\Delta \Phi_n = \frac{\epsilon_n}{M_n} \varphi_n
\label{model1}
\ee
where $\varphi_n$ is the random process on the twosphere given by
\be
\varphi_n = 
\sum_{l\ge2} \varphi_{lmn} Y_{lm}
\label{model1a}
\ee
with
$\langle \varphi_{lmn} \rangle=0$
and
\be
\langle \varphi_{lmn} \varphi_{l'm'n'}^* \rangle = c^2_l
\delta_{ll'} \delta_{mm'} \delta_{nn'}.
\ee
Here 
$c_l$ are dimensionless constants
which are of order unity for $l$ of order unity.  We will take the coefficients $c_l$
to fall off exponentially with $l$ as $l \to \infty$, since this is the behavior of the
transmission coefficients that enter into the amplitudes for emitted
Hawking quanta\footnote{There should also be a factor of $\sim l^{-4}$ in $c_l$
due to the presence of the operator ${\cal D}$ in Eq.\ (\ref{DeltaPhi}).  We neglect this factor
since it is unimportant compared to the exponential factor.}.  Here the Kronecker delta $\delta_{nn'}$ enforces the
fact that successive emitted quanta are uncorrelated,
while the other two Kronecker delta factors ensure isotropy.

\subsection{Results for Poincar\'e charges}
\label{sec:results}

The evolution prescription for the Poincar\'e charges given by Eqs.\ (\ref{e111}) -- (\ref{spinevolve})
is similar to the simple Newtonian model of Sec.\
\ref{sec:Newtonian} above; here, however, it has been derived from the
full BMS kinematics.  There are also some differences from the model
of Sec.\ \ref{sec:Newtonian}, aside from the trivial generalization to three dimensions.
First, there is the evolution equation
(\ref{spinevolve}) for the intrinsic angular momentum, which was not tracked in
Sec.\ \ref{sec:Newtonian}.  We can make order of magnitude estimates
of the terms in Eq.\ (\ref{spinevolve}), using $X_n^i \sim M^2$,
$P_n^i \sim 1$, $\Delta P_n^i \sim \Delta E_n \sim M^{-1}$,
and $\Delta J_n^{ij} \sim {\overline {\Delta J}_n^{0i}} \sim 1$.
The first two terms on the right hand side are of order $\sim 1$ while
the third term is $\sim M^{-1}$ and can be neglected.
The random walk estimate then gives that the fluctuations $\delta
S^{ij}$ in $S^{ij}$
at late times are of order $\sim \sqrt{n} \sim M_i$, where $n \sim M_i^2$ is the
number of steps.  The dimensionless angular momentum parameter is then
of order\footnote{Here we are assuming zero mean spin; if the black
  hole starts with some net spin, this will of course bias the
  evolution and lead to a spin down \cite{PhysRevD.14.3260}.  However
  the fluctuations will be still given by Eq.\ (\ref{deltas}) at
  leading order.}
\be
\frac{\delta S^{ij}}{M^2} \sim \frac{M_i}{M^2}.
\label{deltas}
\ee
Hence the spin fluctuations are unimportant until $M \sim \sqrt{M_i}$,
which occurs after the regime $M \sim M_i^{2/3}$ of interest in this
paper.  A similar conclusion was reached in Appendix C of Ref.\ \cite{2013PhRvD..87h4050N}.

A second difference from the model of Sec.\ \ref{sec:Newtonian} is
the angular momentum term (third term) on the right hand side of Eq.\ (\ref{posnevolve1}) for the
position evolution, which does not appear in the corresponding Eq.\ (\ref{posnevolve}).
This term is of order ${\overline {\Delta J}}_n^{0i}/M_n \sim M^{-1}$,
and is statistically independent of the second term which is $\sim 1$
at late times, so it gives a subdominant contribution to the
displacement fluctuations.

To summarize, we have argued that
our model given by
Eqs.\ (\ref{ss24}) and (\ref{bmsmodel})
does not differ in any essential way
from the simple Newtonian model of Sec.\ \ref{sec:Newtonian}.
Consequently, the late-time fluctuations in the
Poincar\'e charges are still given by Eqs.\ (\ref{lateansfn1}), aside from unimportant
changes in the constant coefficients.

\subsection{Results for super center-of-mass charges}
\label{sec:results1}

Turn now to the super center-of-mass charges.  Combining the definition (\ref{Phidef}) of the potential $\Phi_n$,
the result (\ref{cnans1}) for how the charges are updated at each step,
and the model (\ref{model1}) for the charges carried away by the $n$th quantum,
we obtain for the potential $\Phi_n$ at late times
\be
\Phi_n = \sum_{l\ge 2} \sum_m \, {\hat \varphi_{lmn}} Y_{lm}
\label{Phinresult}
\ee
where ${\hat \varphi}_{lmn} = \sum_{n'\le n} \varphi_{lmn'}$.
Combining this with Eq.\ (\ref{model1}) gives
\begin{eqnarray}
  \langle {\hat \varphi}_{lmn} {\hat \varphi}_{l'm'n}^* \rangle &=& \delta_{ll'} \delta_{mm'} c_l^2 \sum_{n'\le n} \frac{1}{2 M_{n'}^2}
\approx \delta_{ll'} \delta_{mm'} c_l^2 \ln(M_i/M),
\label{Phinresult1}
\end{eqnarray}
where $M = M_n = \sqrt{M_i^2 + 1 - n}$ and we recall that $c_l = O(1)$ for $l = O(1)$.  Hence, for $l$ of order
unity, the fluctuations in each $l,m$ component
${\hat \varphi}_{lmn}$
are of order unity in Planck units (neglecting logarithmic factors), and they fall off exponentially with $l$ for large $l$.
Equivalently, the tensors ${\cal Q}_L$ defined by Eq.\ (\ref{Phiexpand}) have fluctuations of order unity for $l$ of order unity.

Consider now the super center-of-mass charges ${\cal J}_L$ defined by Eq.\ (\ref{eq:stt}). 
For $l \ge 3$ these are simply related to ${\cal Q}_L$ by a factor of the mass, by Eq.\ (\ref{JQrelation}).
Hence we have that the fluctuations are of order
\be
   {\cal J}_L \sim M,
   \label{l3}
\ee
for $l \ge 3$ and $l$ of order unity.  For $l=2$, by contrast, the fluctuations are much larger,
since in the second term in Eq.\ (\ref{JQrelation3}) the orbital
angular momentum is of order $J_{0i} \sim M^3$ at late
times\footnote{This is true both for $u_0=0$ and for $u_0 \sim M_i^3$,
  that is, for both versions of the super center-of-mass charges
  discussed in Appendix \ref{app:choicebasis}.}  and the momentum is of
order $P_i \sim 1$, from Eqs.\ (\ref{earlyans}) and (\ref{eq:cm}), giving
\be
   {\cal J}_{ij} \sim M^2.
\label{l2}
\ee

How large are the fluctuations in the spacetime geometry associated
with the fluctuations (\ref{l3}) and (\ref{l2})?  The fluctuations
(\ref{l3}) correspond to displacements of order unity in Planck
units, and so the fluctuations in the geometry are small.  By
contrast, the displacements associated with the fluctuations (\ref{l2})
are of order $\sim M$, and correspond to macroscopic modifications to the geometry
of order unity.  However, these fluctuations are not independent of
the fluctuations in the center of mass and momentum.  In particular,
the charges $P_\alpha$, $J_{\alpha\beta}$, ${\cal J}_L$, which are in the initial Bondi frame,
determine a BMS transformation to the comoving Bondi frame in which
the metric functions take the simple form (\ref{stat}).    The
$l=2$ supertranslation piece of this transformation is
determined from ${\cal Q}_{ij}$ which has the 
small fluctuations (\ref{Phinresult1}), not from ${\cal J}_{ij}$ which 
has the large fluctuations (\ref{l2}) (see Eqs.\ (\ref{ccc})).  
The macroscopic fluctuations in the geometry are dominated by the
fluctuating boost and fluctuating translation.

\section{Transient effects}
\label{sec:transient}

\subsection{Overview}
\label{sec:transient1}

The calculations so far in this paper have neglected some 
transient effects that are important for the super center-of-mass
fluctuations at high multipole orders.  Specifically, consider the
black hole charges evaluated on the cut ${\cal S}$ of $\scri^+$ given by $u = u_0$ 
in the initial Bondi frame.  We have assumed that each emission event 
impacts $\scri^+$ either completely before ${\cal S}$, or completely
after ${\cal S}$, since we have counted only complete emission events
and not partial events.  In fact, there will be approximately $\sim M_i^2/M$
emission events or outgoing quanta for which the associated stress energy impacts
$\scri^+$ partially before ${\cal S}$, and partially after ${\cal S}$,
whose contributions to the charges have not been correctly accounted
for.  In this section we will estimate the contribution from these
events.

We start by summarizing the results.  We denote by $\Phi_{\rm tr}$ the
``transient'' contribution from the partially counted quanta to the potential $\Phi$ for the
shear tensor at $u=u_0$.  We parameterize the spectrum of angular
fluctuations of $\Phi_{\rm tr}$ by a quantity $d \Phi_{\rm tr}^2 / d
\ln l$
defined so that\footnote{This notation is an alternative to
  that used in Eq.\ (\ref{Phinresult1}).  The two notations are
  related by replacing the right hand side of Eq.\ (\ref{Phinresult1})
  with $ [ l(2l+1)]^{-1} \delta_{ll'} \delta_{mm'} d \Phi^2 / d \ln l$.}
\be
\left< \int d^2 \Omega \,
\Phi_{\rm tr}^2 \right> = \int d \ln l \, \frac{ d \Phi_{\rm tr}^2 }{d \ln l},
\label{spectrumdef}
\ee
where the angular brackets denote expected value.
Here $l$ denotes multipole order, which we treat as a continuous
variable at large $l$.  Our result for the fluctuations of $\Phi_{\rm
  tr}$ at large $l$ is
\be
\frac{ d \Phi_{\rm tr}^2 }{d \ln l} \sim \frac{\sigma_\Delta}{l^9 M^3}
\left( 1 + \frac{M^2 l^2}{2 \sigma_\Delta^2} \right)
\exp \left[ - \frac{l^2 M^2 }{2 \sigma_\Delta^2} \right],
\label{semifinalans}
\ee
where the symbol $\sim$ means that we have dropped constant factors of
order unity.  Here $M$ is the expected mass of the black hole and
$\sigma_\Delta^2$ the variance in the center-of-mass 
location at retarded time $u_0$.  Using the late-time result (\ref{latexansfn1}) for this
variance\footnote{Note that the result (\ref{latexansfn1}) is correct
  up to an unknown constant factor of order unity, arising from the
  idealized model (\ref{simple}),   and hence the argument of
  the exponential factor in Eq.\ (\ref{finalans1}) is similarly subject to a
  correction factor of order unity.} we can rewrite the power spectrum as
\be
\frac{ d \Phi_{\rm tr}^2 }{d \ln l} \sim \frac{M_i^2}{l^9 M^3}
\left( 1 + \frac{2 M^2 l^2}{M_i^4} \right)
\exp \left[ - \frac{2 l^2 M^2 }{M_i^4} \right].
\label{finalans1}
\ee
Thus the transient contribution to the spectrum is a power law up to a critical angular scale
given by $l_{\rm crit} \sim M_i^2 / M$, and at higher $l$ it
is exponentially suppressed.
The transient contribution (\ref{finalans1}) is small compared to the
previously computed contribution (\ref{Phinresult1}) for $l$ of order unity,
but dominates for $1 \ll l \ll l_{\rm crit}$ where the previous
contribution is exponentially suppressed.

The spectrum (\ref{finalans1}) characterizes the fluctuations in the potential
$\Phi$, or equivalently of the tensors ${\cal Q}_L$, at large $l$.
The super center-of-mass charges ${\cal J}_L$ are related to these
tensors by a factor of the black hole mass, as discussed in
Sec.\ \ref{sec:results1} above.

Finally, we note that the shear tensor (\ref{Phidef}) is related to $\Phi$ by
two angular derivatives.  Hence the spectrum of fluctuations of
$C_{AB}$ is given by the right hand side of Eq.\ (\ref{finalans1}) multiplied
by $l^4$, which scales $\propto l^{-5}$.
The total rms fluctuation in $C_{AB}$ is of order
\be
C_{\rm rms}^2 \sim \ln\left(\frac{M_i}{M}\right) + \frac{M_i^2}{ M^3}.
\ee
Here the first term is the contribution (\ref{Phinresult1}) previously
computed, and the second term is the transient contribution
(\ref{finalans1}).  The first term dominates at early times, while the
second term begins to dominate when $M$ becomes small compared to  $M_i^{2/3}$.

\subsection{Derivation}

We now turn to the derivation of the spectrum (\ref{semifinalans}).
Our derivation is based on the same kind of heuristic model as used in
earlier sections of the paper.  A more rigorous derivation based on the
two point function of the flux operator yields qualitatively the same result and
will be given elsewhere.

As previously discussed, each outgoing quantum is characterized by an
outgoing flux $\sim M^{-2}$ over a timescale $\sim M$.  For simplicity
we will assume that the dependence on time is identical for each
outgoing quantum.  Thus for the $n$th quantum we assume,
consistently with Eqs. (\ref{bmsmodel}) and (\ref{model1}), 
\be
{\hat T}_{uu}(u,\theta^A) + {\cal T}_{uu}(u,\theta^A) = M_n^{-2}
\epsilon_n {\cal F}\left[ \frac{ u - u_n  }{M_n} \right] \varphi_n(\theta^A),
\ee
where $u_{n+1} - u_n = M_n$ [cf.\ Eq.\ (\ref{timeevolve})] and ${\cal
  F}$ is a fixed smooth nonnegative function with
${\cal F}(x) = 0$ for $|x| > 1$ and $\int dx {\cal F}(x)=1$.
Also $\varphi_n$ is the random process given by Eq.\ (\ref{model1a}) but now with the $l=0,1$ terms included.
Next, we sum over all the quanta and transform from the instantaneous Bondi frame to the initial
Bondi frame, neglecting the relative boost 
in accordance the slow motion approximation of Sec.\ \ref{sec:slow}.
This gives
\be
{\hat T}_{uu}(u,\theta^A) + {\cal T}_{uu}(u,\theta^A) = 
\sum_n M_n^{-2} \epsilon_n {\cal F}\left[ \frac{ u - u_n + {\bf n} \cdot {\bf \Delta} }{M_n} \right] \varphi_n(\theta^A),
\ee
where ${\bf \Delta} = {\bf \Delta}(u)$ is the center of mass location of the black
hole. Finally we evaluate the net change in the potential $\Phi$ for the shear tensor from
$u=-\infty$ to $u = u_0$ by applying
Eq.\ (\ref{DeltaPhi}) in the initial Bondi frame, neglecting the
subdominant second term, and using Eq.\ (\ref{DeltaE}).  This gives
\be
{\cal D} \Phi(u_0,\theta^A) = 4 \pi {\cal P} \sum_n
M_n^{-1} \epsilon_n \, 
{\hat {\cal F}}\left[ \frac{ u_0 - u_n + {\bf n} \cdot {\bf \Delta} }{M_n}
  \right] \varphi_n(\theta^A).
\label{fa}
\ee
Here we have defined the function ${\hat {\cal F}}(x) =
\int_{-\infty}^x dx' {\cal F}(x')$, which satisfies
\begin{subequations}
  \label{ident1}
  \begin{eqnarray}
   {\hat {\cal F}}(x) &=& 0, \ \ x < -1, \\
   {\hat {\cal F}}(x) &=& 1, \ \ x > 1.
  \end{eqnarray}
\end{subequations}

There are two types of terms that arise in the sum (\ref{fa}).  For
sufficiently early quanta
for which the argument of ${\hat {\cal F}}$ is larger than $1$, we can
drop the ${\hat {\cal F}}$ factor by Eqs.\ (\ref{ident1}), and the computation
reduces to that of Sec.\ \ref{sec:results1}.  
For later quanta for which the argument of ${\hat {\cal F}}$ is less than one in
absolute value, the computation is modified.
These are the terms with $|u_0 - u_n| \lesssim \Delta \sim M^2$, that
is, the final $\sim M$ quanta in the sum.
These terms will enhance
the fluctuations at high multipole orders $l$.  

We now make an order of magnitude estimate the spectrum of
fluctuations as a function of angular scale of the expression
(\ref{fa}).  We specialize to high multipole orders $l$, 
for which the sum over quanta 
will be dominated by the late quanta just discussed.  For these terms we can drop the factor
$\varphi_n(\theta^A)$, since its dependence on $\theta^A$ is
exponentially small at large $l$, and we have $\varphi_n \sim 1$ at low $l$.
We also approximate ${\bf \Delta}(u)$ by its final value ${\bf
  \Delta}(u_0)$, and $M_n$ by its final value at $u = u_0$ which we
denote simply by $M$.
We assume for simplicity that there is a value ${\bar n}$ of $n$ for
which $u_{\bar n} = u_0$, and define $j = n - {\bar n}$, so that to a good
approximation we have $u_n = u_0 + j M$.
We assume initially that ${\bf \Delta}$ is fixed with $\Delta = | {\bf \Delta}
| \gg M$; later we will consider the effect of fluctuations in ${\bf \Delta}$.
We will also approximate $\Delta/M$ by the integer $l_*$ that is
closest to it,
\be
l_* = \left[ \frac{\Delta}{M} \right],
\label{lstardef}
\ee
and define $\mu$ to be the cosine of
the angle between ${\bf n}$ and ${\bf \Delta}$.  With these
definitions and approximations we have that the relevant terms in
Eq.\ (\ref{fa}) are 
\be
{\cal D} \Phi_{\rm tr}(u_0,\theta^A) = \frac{4 \pi}{M} {\cal P} \sum_{j=-l_*}^{l_*}
 \epsilon_{{\bar n} + j} \, 
{\hat {\cal F}}(l_* \mu - j),
\label{fa1}
\ee
where the subscript ``tr'' denotes the transient contribution to $\Phi$
from the late incomplete quanta.

We can divide up the sphere into $2 l_*$ strips of width $\sim 1/l_*$,
with the $k$th strip given by $\mu_k \le \mu \le \mu_{k+1}$ for $-l_*
\le k < l_*$, where $\mu_k = k/l_*$.  On the $k$th strip the terms
in the sum (\ref{fa1}) with $j > k+1$ vanish, while the terms with $j
< k$ are constant, by Eqs.\ (\ref{ident1}), which yields
\be
{\cal D} \Phi_{\rm tr}(u_0,\theta^A) = \frac{4 \pi}{M} {\cal P} \left\{ \sum_{j=-l_*}^{k-1}
 \epsilon_{{\bar n} + j}  + \epsilon_{{\bar n} + k} {\hat {\cal
     F}}[l_*(\mu - \mu_k)] + \epsilon_{{\bar n} + k+1} {\hat {\cal
     F}}[l_*(\mu - \mu_{k+1})] \right\}. 
\label{fa2}
\ee
We now estimate the total power in the fluctuations by squaring and
integrating over the two sphere, which reduces to a sum over strips of
the integral over each strip.  In this calculation we drop the second
and third terms in the brackets in Eq.\ (\ref{fa2}), thereby making a
fractional error of order $1/l_* \ll 1$.
The projection operator ${\cal P}$ subtracts off
$l=0,1$ modes, which has the effect of
replacing $\epsilon_{{\bar n} + j}$ with $\delta \epsilon_{{\bar n} +
  j} = \epsilon_{{\bar n}+j} - \langle \epsilon_{{\bar n}+j} \rangle$.
It also changes the final answer by a factor of two which we will neglect.
We obtain
\be
\left< \int d^2 \Omega  ({\cal D} \Phi_{\rm tr})^2 \right> \sim \frac{1}{M^2 l_*}
\sum_{k=-l_*}^{l*-1} \left< \left( \sum_{j=-l_*}^{k-1} \delta \epsilon_{{\bar n}+j} \right)^2 \right>
\sim \frac{1}{M^2 l_*} \sum_k (k+l_*) \sim \frac{l_*}{M^2},
\label{ansg}
\ee
where the angular brackets denote expected value and we have used Eq.\ (\ref{vareps}).

Now the function $\sum_j \delta \epsilon_{{\bar n} + j}$ describes a random walk,
and hence the spectrum $d ({\cal D} \Phi_{\rm tr})^2 / d \ln l$ of the fluctuations scales $\propto 1/l$
where we are using the notation (\ref{spectrumdef}),
since this is a well-known property of random walks
\cite{2017mcp..book.....T}.  This powerlaw spectrum continues up to
the maximum scale $l \sim l_*$ of the individual strips.   
Combining this with the normalization (\ref{ansg}) and the definition
(\ref{spectrumdef}) (with $\Phi_{\rm tr}$ replaced by ${\cal D}
\Phi_{\rm tr}$) yields
\be
\frac{ d ({\cal D} \Phi_{\rm tr})^2} {d \ln l} \sim \frac{l_*}{M^2 l}
\Theta(l_* - l) \Theta(2 - l),
\ee
where $\Theta$ is the step function.
Using ${\cal D} \sim l^4$ from Eq.\ (\ref{calDdef}) it follows that
\be
\label{almost}
\frac{ d  \Phi_{\rm tr}^2} {d \ln l} \sim \frac{l_*}{M^2 l^9}
\Theta(l_* - l) \Theta(2 - l).
\ee

So far in this discussion we have treated ${\bf \Delta}$ as fixed.  We now
take into account that ${\bf \Delta}$ has a distribution that is very
nearly Gaussian, by the central limit theorem, since it is a sum of a large
number of independent contributions (cf.\ Sec.\ \ref{sec:earlytime} above):
\be
\frac{d {\cal P} }{ d \ln \Delta} \sim \frac{\Delta^3}{\sigma_\Delta^3} \exp \left[ - \frac{ \Delta^2}{2 \sigma_\Delta^2} \right],
\ee
where $\sigma_\Delta^2$ is the variance.
We can now integrate this against the expression (\ref{almost}) to get the total spectrum:
\be
\frac{ d \Phi_{\rm tr}^2} {d \ln l}(l) = \int d \ln \Delta \ \frac{ d
  \Phi_{\rm tr}^2} {d \ln l}(l;\Delta)  \  \frac{d {\cal P} }{ d \ln
  \Delta}(\Delta),
\ee
which using Eq.\ (\ref{lstardef}) yields the
final result (\ref{semifinalans}).

\acknowledgments

I thank Abhay Ashtekar, Venkatessa Chandrasekharan and Kartik Prabhu
for helpful discussions.  This research was supported in part by NSF
grants PHY-1404105 and PHY-1707800.

\appendix

\section{Derivation of late time predictions of stochastic process}
\label{app:derive} 

In this appendix we derive the late time predictions (\ref{lateans})
of the Newtonian stochastic model (\ref{simple}) of black hole evolution of Sec.\ \ref{sec:Newtonian}.

We start by defining
\begin{subequations}
\label{decompose}
\begin{eqnarray}
\label{massdecompose}
{\bar M}_n &= \langle M_n \rangle, \ \ \ \ \ \delta M_n &= M_n - {\bar
  M}_n, \\
{\bar t}_n &= \langle t_n \rangle, \ \ \ \ \ \delta t_n &= t_n - {\bar
  t}_n,
\end{eqnarray}
\end{subequations}
where the angular brackets denote an expectation value.
Substituting the decomposition (\ref{massdecompose}) into the mass
evolution equation
(\ref{massevolve}),
expanding in powers of $\delta M_n$, and separating the expected value
and the remaining part of the equation gives
\begin{subequations}
\label{sevolve}
\begin{eqnarray}
\label{sevolvem}
{\bar M}_{n+1} &=& {\bar M}_n - \frac{1}{2 {\bar M}_n} \left[ 1 + O \left(
    \frac{\delta M_n^2}{{\bar M}_n^2} \right) \right], \\
\label{sevolvedm}
\delta M_{n+1} &=& \delta M_n + \left( \frac{\delta M_n}{2 {\bar M}_n^2}
- \frac{\delta \epsilon_n}{{\bar M}_n} \right)
\left[ 1 + O \left(
    \frac{\delta M_n}{{\bar M}_n} \right) \right],
\end{eqnarray}
\end{subequations}
where $\delta \epsilon_n = \epsilon_n - \left< \epsilon_n \right>$.
The first equation gives just the usual semiclassical evolution of the
expected mass of the black hole, while the second gives the evolution
of the mass fluctuations.

The solution for the expected mass can be obtained by approximating
Eq.\ (\ref{sevolvem}) as a differential equation, and is\footnote{The
  error estimate can be obtained by computing the solution to
  subleading order, which is ${\bar M}_n = Q_n - \ln(Q_n/M_i)/(4 Q_n)$ with $Q_n^2
  = M_i^2 + 1 - n$.}
\be
{\bar M}_n = \sqrt{M_i^2 + 1 - n} \left[ 1 + O \left( \frac{M_i -
      M}{M_i M^2} \right) \right].
\label{barMans}
\ee
Here on the right hand side we have written $M$ for ${\bar M}_n$
inside the error estimates, for simplicity.  We have also temporarily
dropped the error term $\delta M_n^2/{\bar M}_n^2$ from Eq.\
(\ref{sevolvem}); we will restore this fractional error estimate at
the end of
the computation, when we have computed $\langle \delta M_n^2 \rangle$.
Next from the expected value of the time evolution equation
(\ref{timeevolve})
and the definitions (\ref{decompose})
we find
${\bar t}_{n+1} = \sum_{k=1}^n {\bar M}_k$.  Converting this to an
integral\footnote{We use the approximation
$$
\sum_{k=a}^b f(k)  = \int_{a-\frac{1}{2}}^{b+\frac{1}{2}} dk f(k) \left[1 + O\left(
    \frac{f''}{f} \right) \right].
$$
}
and using the expression (\ref{barMans}) gives
\be
{\bar t}_n = \frac{2}{3} (M_i^3 - {\bar M}_{n-1}^3) \left[ 1 +
O\left( \frac{1}{M_i^2} \right)
+O\left( \frac{1}{M^4} \right)
\right].
\label{bartans}
\ee

We now turn to computing the fluctuations.
From Eq.\ (\ref{sevolvedm}) we obtain
\begin{eqnarray}
\delta M_{n+1} &=& - \sum_{j=1}^n \left[\prod_{k=j+1}^n \left( 1 +
    \frac{1}{2 {\bar M}_k^2} \right) \right] \frac{\delta
  \epsilon_j}{{\bar M}_j}
\left[ 1 + O \left( \frac{{\delta M}_n}{{\bar M}_n} \right) \right]
\label{deltam1}
\end{eqnarray}
The product inside the square brackets can be
evaluated by taking the logarithm,
converting the sum to an integral,
and using Eq.\ (\ref{barMans}),
which yields
\be
\delta M_{n+1} = - \frac{q_{n+1}}{{\bar M}_n} \left[ 1 + O({\bar
    M}_n^{-2}) +
O \left( \frac{{\delta M}_n}{{\bar M}_n} \right) \right],
\label{deltam2}
\ee
where $q_{n+1} = \sum_{k=1}^n \delta \epsilon_j$.
Squaring and taking the expected value gives $\langle \delta M_{n+1}^2
\rangle = n/(4 {\bar M}_n^2)$.  Using this expression to evaluate
the error estimate in Eq.\ (\ref{deltam2})  and eliminating $n$ in
favor of $M = {\bar M}_n$ using (\ref{barMans})
finally yields
\be
\label{latemansfn}
\langle \delta M_n^2 \rangle =  \frac{M_i^2 - M^2}{4 M^2} \left[1 + O\left(
\frac{\sqrt{M_i^2-M^2}}{M^2}\right)\right].\\
\ee
Note that this is the fluctuation in mass at fixed $n$, to be
distinguished from the more physically relevant fluctuations in mass at
fixed time $t$ [cf.\ Eq.\ (\ref{latemans}) above], which we compute
below.

We next compute the fluctuations $\delta t_n$.  From the time
evolution equation (\ref{timeevolve}) we obtain $\delta t_{n+1} =
\sum_{k=1}^n \delta M_k$, and squaring and taking the expected value
using Eq.\ (\ref{deltam2}) gives
\begin{eqnarray}
\langle \delta t_{n+1}^2 \rangle &=& \sum_{k,l=1}^n \frac{ {\rm
    min}(k-1,l-1)}{4 {\bar M}_k {\bar M}_l}
\left[ 1 + O \left( \frac{\sqrt{k}}{{\bar M}_k} \right)
+ O \left( \frac{\sqrt{l}}{{\bar M}_l} \right) \right].
\end{eqnarray}
Converting the sums to integrals as before yields
\begin{eqnarray}
\langle \delta t_{n+1}^2 \rangle &=&
\frac{1}{2} (M_i-M)^3 (M_i + M/3)
\left[ 1 + O \left( \frac{\sqrt{M_i^2 - M^2}}{M^2} \right)\right].
\label{latetansfn}
\end{eqnarray}

Now in this simple discrete model of the black hole evolution,
the black hole mass $M(t)$ at a given time $t$ is obtained
by evaluating $M_n$ at the value of $n$ for which $t_n$ is closest to
$t$.  Hence the fluctuations in mass at fixed time $t$ are given by
\be
\delta M(t) = \delta M_n - M'(t) \delta t_n,
\label{deltaMft}
\ee
where the right hand side is evaluated at the value of $n = n(t)$
obtained by solving Eqs.\ (\ref{barMans}) and (\ref{bartans}),
and the derivative is given by $M'(t) = -1/(2 M^2)$.
Squaring Eq.\ (\ref{deltaMft}), taking the expected value, dropping
the cross term which one can show is subdominant, and using the
expressions (\ref{latemansfn}) and (\ref{latetansfn}) finally yields
the result (\ref{latemans}).  Note that the mass fluctuations
(\ref{latemans}) at fixed
time are dominated by the fluctuations in $t_n$.

To compute the momentum fluctuations, we expand Eq.\ (\ref{momevolve})
in powers of $\delta M_n$ and solve, obtaining
\be
p_{n+1} = \sum_{k=1}^n \frac{\epsilon_k \delta_k}{{\bar M}_k}
\left[ 1
  - \frac{{\delta M}_k}{{\bar M}_k}
+ O \left( \frac{ {\delta M}_k^2 }{ {\bar M}_k^2 } \right)
\right].
\label{pnans}
\ee
Squaring and taking the expected value gives
\be
\langle p_{n+1}^2 \rangle = \sum_{k=1}^n \frac{1}{2 {\bar M}_k^2}
\left[ 1
+ O \left( \frac{ {\delta M}_k^2 }{ {\bar M}_k^2 } \right)
\right],
\ee
since $\delta M_k$, $\epsilon_k$ and $\delta_k$ are statistically
independent.  Converting the sum to an integral and using Eqs.\
(\ref{barMans}) and (\ref{latemansfn}) now yields the formula
(\ref{latepans}).

Finally, by combining the position evolution equation
(\ref{posnevolve}) with the expression (\ref{pnans}) for the momentum
yields
\be
 x_{n+1}^2 = \sum_{r,s=1}^n \sum_{k=1}^{r-1}
\sum_{l=1}^{s-1} \frac{\epsilon_k \delta_k \epsilon_l \delta_l}{{\bar
    M}_k {\bar M}_l} \left[1 + O \left( \frac{ \delta M}{{\bar M}}
  \right) \right].
\ee
Taking the expected value and converting the sums to integrals gives
\begin{eqnarray}
\langle x_{n+1}^2 \rangle&=& \frac{1}{2} \sum_{r,s=1}^n \sum_{k=1}^{{\rm
    min}(r,s)-1}
\frac{1}{{\bar
    M}_k^2}  \left[1 + O \left( \frac{ \delta M}{{\bar M}}
  \right) \right]
\nonumber \\
&=& 4 \int_{M_n}^{M_i} d M_r \int_{M_n}^{M_i} d M_s
\int_{{\rm max}(M_r,M_s)}^{M_i} dM_k
\frac{M_r M_s}{M_k}
 \left[1 + O \left( \frac{ \delta M}{{\bar M}}
  \right) \right],
\end{eqnarray}
and evaluating the integral gives the expression (\ref{latexans}).

\section{Independent charges in stationary regions of future null infinity}
\label{app:indepcharges}

In this appendix we derive the relationships between the various charges
of the extended BMS algebra that apply in stationary regions of future null infinity,
and show that the independent charges can be taken to be the
4-momentum $P^\alpha$, the angular momentum $J^{\alpha\beta}$, and the
super center-of-mass charges.  Equivalently, we show that the
supermomentum and superspin charges are determined in terms of the
other charges, and so can be neglected for our purposes.

In the canonical Bondi frame associated with the stationary region,
the metric functions take the simple form (\ref{stat}).  We now make a
nonlinear BMS transformation to a general Bondi frame, of the form (FN,2.12), following
Appendix B of FN.  Quantities in the new frame will be
denoted with overbars.  The transformation is parameterized in
terms of a conformal isometry $\varphi : S^2 \to S^2$ of the 2-sphere into itself,
and a function $\beta$ on the two sphere [denoted by $\alpha$ in
Eq.\ (FN,2.12)].  The Bondi mass aspect in this general frame is given
from Eqs.\ (FN,B5) and (\ref{stat}) as
\be
   {\bar m}(\theta^A) = m_0 \omega_\varphi^{-3},
   \label{bondi11}
\ee
where $\omega_\varphi(\theta^A)$ is defined by $\varphi_* h_{AB} =
\omega_\varphi^{-2} h_{AB}$ and $\varphi_*$ is the pullback. The
quantity $\omega_\varphi$ is determined by the boost part of the
Lorentz transformation $\varphi$ and is given explicitly by
\be
\omega_\varphi = \cosh \psi - {\bf n} \cdot {\bf m} \sinh \psi,
\label{omegaformula}
\ee
where ${\bf n} = (\sin\theta \cos\phi, \sin \theta \sin\phi,
\cos\theta)$, $\psi$
is the rapidity parameter of the boost and ${\bf m}$ is
a unit vector giving the direction of the velocity of the general
frame with respect to the canonical frame.  The 4-momentum in the
general frame is now
given from Eqs.\ (\ref{charge-general}), (FN,3.7) and (FN,3.9) as
\be
{\bar P}^\alpha = ({\bar P}^0,{\bf {\bar P}})  = \frac{1}{4 \pi} \int d^2 \Omega (1, {\bf
  n}) m_0 \omega_\varphi^{-3} = m_0 (\cosh \psi,  \sinh \psi {\bf
  m}).
\label{barPP}
\ee
We therefore see that that the Bondi mass aspect (\ref{bondi11}) and
all the supermomentum charges are determined in terms of the Bondi
4-momentum (\ref{barPP}).

Turn now to the superspin and super center-of-mass charges, which are
encoded in the function ${\hat N}_A$ defined by
Eq.\ (\ref{hatNdef}). The transformation law for this function
in stationary regions of $\scri^+$ can be obtained by combining Eqs.\ (FN,B1), (FN,B2)
and (\ref{charge-general}) together with footnote 25 of
FN and is
\be
   {\bar {\hat N}}_A = \omega_\varphi^{-2} \varphi_* {\hat N}_A + 3
   \omega_\varphi^{-3} (\varphi_* m ) D_A \beta + \beta D_A (
   \omega_\varphi^{-3} \varphi_* m ).
\label{tft}
   \ee
Here the overbar denotes the value of this function in the general
Bondi frame.  The transformation law (\ref{tft}) can be simplified by
defining the new quantity
\be
S_A = {\hat N}_A + \frac{3}{2} m D_A \Phi + \frac{1}{2} \Phi D_A m,
\label{SAdef}
\ee
where the potential $\Phi$ for the electric parity piece of the shear
tensor $C_{AB}$ is defined in Eq.\ (\ref{Phidef}).
Now in the canonical BMS frame, $S_A$ will coincide with ${\hat N}_A$
and will be purely $l=1$ and of magnetic parity, encoding the
intrinsic spin of the spacetime.
Combining Eqs.\ (\ref{tft}), (\ref{SAdef}), (FN,B5), (FN,B6) and
(FN,B8) yields that in the general BMS frame this function will be
\be
   {\bar S}_A = \omega_\varphi^{-2} \varphi_* S_A + 3 {\bar m} D_A \beta_0 + \beta_0 D_A {\bar m},
   \label{barSdef}
\ee
where $\beta_0$ consists of the $l=0,1$ components of $\beta$.
Combining this with a barred version of Eq.\ (\ref{SAdef})
gives
\be
   {\bar {\hat N}}_A =
-\frac{3}{2} {\bar m} D_A {\bar \Phi} - \frac{1}{2} {\bar \Phi} D_A
  {\bar m}
   + \omega_\varphi^{-2} \varphi_* S_A + 3 {\bar m} D_A \beta_0 + \beta_0 D_A {\bar m}.
   \label{scom}
   \ee
The last three terms in this equation are 
determined from the intrinsic spin, 4-momentum 
and from the Poincar\'e transformation $(\varphi,\beta_0)$ relating
the two frames, so they are determined by the linear and angular momentum ${\bar P}^\alpha$ and ${\bar J}^{\alpha\beta}$. 
It follows that ${\bar {\hat N}}_A$ and the superspin and super
center-of-mass charges are determined from
${\bar J}^{\alpha\beta}$ and ${\bar P}^\alpha$ and the electric parity potential
${\bar \Phi}$, in a general Bondi frame in a stationary region\footnote{This
  result was previously derived within a limited approximation in
  Sec.\ III.E of FN.}.

We next derive the explicit form of the super center-of-mass charges,
expanding to second order in the velocity of the boost.
Since the metric is stationary, shifting the supertranslation $\beta$ by a
constant times $\omega_\varphi$ does not affect the metric in the
general Bondi frame, from Eqs.\ (FN,2.12).  Therefore without loss of generality we take
the $l=0$ component of $\beta$ to vanish, and we parameterize the
translation $\beta_0$ as $\beta_0 = \beta_i n^i$.
From Eq.\ (\ref{scom}) we can now read off 
the orbital angular momentum (\ref{J0idef}) and the $l=2$ super
center-of-mass charge (\ref{superboost}) in the general frame,
using Eqs.\ (\ref{bondi11}) and (\ref{omegaformula}) to expand to
expand in powers of the velocity ${\bf v} ={\bf m} \tanh \psi$ of the boost:
\begin{subequations}
  \label{ccc}
\begin{eqnarray}
  {\bar J}_{0i} &=& - J_{ij} v^j + m_0 \left(1 + \frac{1}{2}
  v^2\right) \beta_i - \frac{4}{5} m_0 {\bar {\cal Q}}_{ij} v^j,\\
  {\bar {\cal J}}_{ij} &=& - \frac{3}{5} m_0 {\bar {\cal Q}}_{ij}
  \left[ 1 + O(v^2) \right] + \frac{8}{5} v_{<i} J_{j>k} v^k +
  \frac{12}{5} m_0 \beta_{<i} v_{j>}.
  \label{JQrelation1}
\end{eqnarray}
\end{subequations}
Here we have used the definition (\ref{Phiexpand}),
$J_{ij}$ is the
intrinsic angular momentum in the canonical Bondi frame, and the
angular brackets $< \ldots >$ denote the symmetric tracefree projection.
Combining Eqs.\ (\ref{ccc}) to eliminate $\beta_i$, dropping the
intrinsic angular momentum terms and using Eq.\ (\ref{barPP}) gives
\be
  {\bar {\cal J}}_{ij} = - \frac{3}{5} m_0 {\bar {\cal Q}}_{ij}
  \left[ 1 + O(v^2) \right] +
  \frac{12}{5 m_0}  {\bar J}_{0<i} {\bar P}_{j>} \left[ 1 + O(v^2) \right].
  \label{JQrelation2}
\ee
A similar calculation for $l \ge 3$ using Eq.\ (\ref{eq:stt}) gives
\be
{\bar {\cal J}}_L = - \frac{3 m_0}{2 g_l} {\bar {\cal Q}}_L  + O(v^2).
\label{JQrelation}
\ee

\section{Choice of basis of algebra of charges}
\label{app:choicebasis}

In this appendix we discuss two different versions of the super
center-of-mass charges.  To explain these versions,
it is useful to distinguish between two different kinds of time
evolution of the charges.  The first is just the kind discussed in
Sec.\ \ref{sec:transition}, associated with evaluating the charges as surface
integrals on cuts of $\scri^+$ of the form $u = $ constant, and
varying $u$.

A second kind of time evolution is associated with the
choice of basis in the algebra of asymptotic symmetries.
Consider for example the orbital angular momentum $J_{0i}$ that is associated
via Eqs.\ (\ref{bmsinfinitesimal}) and (\ref{charge-general}) 
with the boost symmetry generator ${\vec \xi} = D^A n^i \partial_A - u n^i \partial_u$.
This choice of boost symmetry is associated with a particular choice
of origin of the retarded time coordinate $u$ (which in the body of the paper we took to
be the time of formation of the black hole, see 
Sec.\ \ref{sec:evolution}).  However, by conjugating the symmetry
generator with a time translation $ u \to u - u_0$ where $u_0$ is a
constant, we can obtain the new boost symmetry generator
\be
   {\vec \xi} = D^A n^i \partial_A - (u-u_0) n^i \partial_u.
\ee
We denote the corresponding charge by $J_{0i}(u,u_0)$, where the first
argument reflects the dependence on the cut of $\scri^+$ and the
second argument the choice of generator.  The dependence on $u_0$
is given by
\be
J_{0i}(u,u_0) = J_{0i}(u,0) + u_0 P_i(u);
\label{eq:gc}
\ee
changing $u_0$ amounts to a change of basis in the algebra of symmetry
generators or equivalently in the algebra of charges.  The
center of mass at retarded time $u$, given by
\be
X_i(u) = \frac{1}{P^0(u)} J_{0i}(u,u)
= \frac{1}{P^0(u)} \left[ J_{0i}(u,0) + u P_i(u) \right],
\label{eq:cm}
\ee
encodes both types of time dependence.
It is this quantity and not the charge $J_{0i}(u,0)$ that most
directly enters into the metric at retarded time $u$, from
Eqs.\ (\ref{metric}), (\ref{NAdef}), (\ref{charge-general}) and (\ref{hatNdef}).


There is an exactly analogous story for the super center-of-mass
charges \cite{Compere:2019gft}.
We specialize for simplicity to the quadrupole $l=2$ case.
Consider the superboost symmetry generator given by (\ref{bmsinfinitesimal}) for $\alpha = 0$, $Y_A =
D_A(n_i n_j)$,
\be
{\vec \xi} = D^A(n_i n_j- \delta_{ij}/3) \partial_A - 3 u (n_i n_j - \delta_{ij}/3)
\partial_u.
\label{superboost}
\ee
We denote the corresponding charge (\ref{charge-general}) by ${\cal J}_{ij}(u)$, a symmetric
traceless tensor
[cf.\ the discussion around Eq.\ (\ref{eq:stt}) above].  As before by conjugating with a time translation we
can obtain a new superboost symmetry generator, given by
Eq.\ (\ref{superboost}) with $u$ replaced by $u - u_0$, and we denote
the corresponding charge by ${\cal J}_{ij}(u,u_0)$.  The dependence on
$u_0$ is given by, from Eqs.\ (\ref{charge-general}) and
(\ref{superboost}),
\be
{\cal J}_{ij}(u,u_0) = {\cal J}_{ij}(u,0) + 3 u_0 {\cal P}_{ij}(u),
\label{eq:gc1}
\ee
where
\be
{\cal P}_{ij} = \frac{1}{4 \pi} \int d^2 \Omega m \left(n_i n_i -
\frac{1}{3} \delta_{ij}\right)
\label{calPdef}
\ee
is the $l=2$ supermomentum charge.  As before we can define a super
center-of-mass quantity that incorporates both types of time
evolution, and which is the quantity that appears most directly in the
metric, via
\be
X_{ij}(u) = \frac{1}{P^0(u)} {\cal J}_{ij}(u,u)
= \frac{1}{P^0(u)} \left[ {\cal J}_{ij}(u,0) + 3 u {\cal P}_{ij}(u) \right].
\label{eq:cm1}
\ee
We will call the charge (\ref{eq:gc1}) the super
center-of-mass charge, and the quantity (\ref{eq:cm1}) the {\it comoving}
super center-of-mass, following Comp\`ere \cite{Compere:2019gft}.

In the special case of stationary regions of $\scri^+$ the charges
${\cal J}_{ij}(u,u_0)$ and $J_{0i}(u,u_0)$ are related by the
formula (\ref{JQrelation2}) derived in Appendix
\ref{app:indepcharges}:
\be
  {\cal J}_{ij}(u,u_0) = - \frac{3}{5} m_0 {\cal Q}_{ij}(u)
  \left[ 1 + O(v^2) \right] +
  \frac{12}{5 m_0}  J_{0<i}(u,u_0) P_{j>}(u)  \left[ 1 + O(v^2) \right].
  \label{JQrelation3}
\ee
Here $m_0$ is the rest mass associated with the Bondi 4-momentum,
$v$ is the velocity of the Bondi frame with respect to the canonical
Bondi frame, ${\cal Q}_{ij}$ is defined
in Eq.\ (\ref{Phiexpand}), and the angular brackets denote symmetric
tracefree projection.  This formula is consistent with the
transformation laws (\ref{eq:gc}) and (\ref{eq:gc1}) because 
we have in stationary regions
\be
   {\cal P}_{ij} = \frac{4}{5 m_0} \left(P_i P_j - \frac{1}{3} P^2
   \delta_{ij} \right) \left[ 1 + O \left( \frac{P^2}{m_0^2} \right) \right],
   \label{calPij}
   \ee
from Eqs.\ (\ref{calPdef}), (\ref{bondi11}) and (\ref{omegaformula}).
Now evaluating Eq.\ (\ref{JQrelation3}) at $u=u_0$, dividing by
$P^0(u)$ and using the definitions (\ref{eq:cm}) and (\ref{eq:cm1}) gives
the relation between the comoving super center-of-mass and normal
center of mass
\be
  X_{ij}(u) = - \frac{3}{5} {\cal Q}_{ij}(u)
  \left[ 1 + O(v^2) \right] +
  \frac{12}{5 m_0}  X_{<i}(u) P_{j>}(u)  \left[ 1 + O(v^2) \right].
  \label{COMrelation}
\ee

\section{Derivation of changes in charges in stationary-to-stationary transitions}
\label{app:fluxes}

In this appendix we derive the formulae
(\ref{Delta4P}), (\ref{DeltaJs}) and (\ref{DeltaPhi})
for the changes in BMS and extended BMS charges in
stationary to stationary transitions in terms of fluxes to null infinity
of mass-energy or gravitational-wave energy.
A similar analysis in a different notation can be found in Ref.\ \cite{Compere:2019gft}.

The change (\ref{Delta4P}) in Bondi 4-momentum is obtained by
multiplying Eq.\ (FN,4.3) by $(1,{\bf n})$, integrating over solid
angles, and noting that the $l=0,1$ components of $\Phi$ vanish by definition.

For general superspin and super center-of-mass charges, we denote by
$Q_Y$ the charge (\ref{charge-general}) specialized to $\alpha=0.$  To
derive the change in this charge we first define the subleading
memory observables
\begin{subequations}
\label{submem}
\begin{eqnarray}
{\widetilde {\Delta \Phi}} &=& \int_{u_1}^{u_2} du \, u \partial_u \Phi,
\\
{\widetilde {\Delta \Psi}} &=& \int_{u_1}^{u_2} du \, u \partial_u \Psi.
\end{eqnarray}
\end{subequations}
These observables\footnote{The electric parity piece
  ${\widetilde {\Delta \Phi}}$
  is called center-of-mass memory
\cite{Nichols:2018qac},
while the magnetic parity piece   ${\widetilde {\Delta \Psi}}$
is called spin memory \cite{Pasterski:2015tva}.}
parameterize the relative displacement, produced by a burst of gravitational waves,
of two test masses that are initially co-located with an initial
relative velocity; see, e.g, Sec.\ II.A of
Ref.\ \cite{Flanagan:2018yzh}.
Now differentiating Eq.\ (\ref{hatNdef}) with respect to $u$ and combining
with Eqs.\ (FN,2.7) and (FN,2.11) gives
\begin{eqnarray}
\partial_u {\hat N}_A &=& - 8 \pi {\hat T}_{uA} + 4 \pi u D_A {\hat
    T}_{uu} + \frac{u}{8} D_A(N_{BC} N^{BC})
  - \frac{3}{8} N_{AB} D_C C^{BC} + \frac{3}{8} C_{AB} D_C N^{BC}
\nonumber \\ &&
+ \frac{1}{8} D_B C_{AC} N^{BC}
- \frac{1}{8} D_B N_{AC} C^{BC}
- \frac{u}{4} D_B D_A D_C N^{BC} + \frac{u}{4} D_B D^B D^C N_{AC}
\nonumber \\
&&
- \frac{u}{4} D_A D_B D_C N^{BC} - 2 \pi \partial_u {\hat T}_{rA}.
\label{hatNAdot}
\end{eqnarray}
We now multiply by $Y^A$ and integrate over solid angles and over
$u$.  The left hand side then becomes the change $\Delta Q_Y$ in the
charge, from Eq.\ (\ref{charge-general}).  On the right hand side, the
last term gives a vanishing contribution since we assume the stress
energy tensor vanishes at $u=u_1$ and $u=u_2$.  The second and third
terms can  be simplified using the definition (\ref{tildeDeltaE}) of $u$-weighted energy
flux ${\widetilde {\Delta {\cal E}}}$, while the first, fourth, fifth, sixth and seventh terms can
be similarly simplified using the definition (\ref{DeltaEA}) of
angular momentum flux $\Delta {\cal E}_A$, making use of Eq.\ (FN,3.23).
The eighth, ninth and tenth terms can be written in terms of the
subleading memory observables (\ref{submem}) using Eqs.\ (\ref{news})
and (\ref{Phidef}), and using $R_{ABCD} = h_{AC} h_{BD} - h_{AD} h_{BC}$.
The final result is\footnote{In the language of
  Ref.\ \cite{Strominger:2017zoo}, the limit $u_1 \to
    -\infty$ and $u_2 \to \infty$ of Eq.\ (\ref{DeltaYans}) expresses
    the total charge on the left hand side as the sum of a hard
    (first) term and a soft (second) term.  In the language of
    Ref.\ \cite{Bieri:2013ada},
    Eq.\ (\ref{DeltaYans}) expresses the total memory (second term) in
    terms of a null piece (first term) and an ordinary piece (left
    hand side) \cite{Nichols:2018qac}.}
\begin{eqnarray}
\Delta Q_Y &=& - \int d^2 \Omega \left[  \Delta {\cal E}_A Y^A  + \frac{1}{2} 
  {\widetilde {\Delta {\cal E}}}
  D_A Y^A  \right] \nonumber \\
&&- \frac{1}{8 \pi} \int d^2 \Omega \left[ {\widetilde {\Delta \Phi}}
  {\cal D} (D_A Y^A) +
  {\widetilde {\Delta \Psi}}
  {\cal D} (\epsilon^{AB} D_A Y_B)  \right],
\label{DeltaYans}
\end{eqnarray}
where the differential operator ${\cal D}$ is given by Eq.\ (\ref{calDdef}).
The changes (\ref{DeltaJs})
in angular momentum components can
now be obtained by taking $Y^A = 2 e^A_{[i} n_{j]}$ 
and $Y^A = e^A_i$, using $D_A e^A_i = D^2 n_i = - 2 n_i$ and the fact
that the operator ${\cal D}$ annihilates the $l=0,1$ components of functions on the sphere,
making the second term in Eq.\ (\ref{DeltaYans}) vanish.

Finally we turn to deriving the change (\ref{DeltaPhi}) in the electric parity
potential $\Phi$ for the shear tensor $C_{AB}$, which encodes the
super center-of-mass charges.
We multiply Eq.\ (FN,4.3) by the projection operator ${\cal
  P}$ that sets to zero the $l=0,1$ pieces of functions on the
sphere, and use ${\cal P} {\cal D} = {\cal D}$ to obtain
\be
{\cal D} \Delta \Phi = 4 \pi {\cal P} \Delta {\cal E} + {\cal P} \Delta m.
\label{ans11}
\ee
To evaluate $\Delta m$ we now specialize to an initial rest Bondi frame in which the spatial
components of the Bondi 4-momentum vanish, so that the initial Bondi
mass aspect is a constant $m_0$, from Eqs.\ (\ref{bondi11}) and (\ref{barPP}).
The final Bondi mass aspect will be of the form (\ref{bondi11}) with
$m_0$ replaced by $m_0 + \delta m$ for some $\delta m$, for some boost
parameters $\psi$ and ${\bf m}$.  Eliminating the boost parameters in
terms of the radiated spatial momentum
using Eqs.\ (\ref{omegaformula}) and (\ref{barPP})
and expanding to second order
in this momentum now gives the formula (\ref{DeltaPhi}).

\bibliographystyle{JHEP}

\providecommand{\href}[2]{#2}\begingroup\raggedright\endgroup

\end{document}